\documentclass[twocolumn,10pt,twoside]{IEEEtran}

\usepackage{amsmath,amsthm,graphicx,bm}
\usepackage{epsfig,amsfonts}
\usepackage{graphicx}
\usepackage{color}
\usepackage{xcolor}

\usepackage[nolist]{acronym}
\usepackage{psfrag}
\usepackage{perso}
\usepackage{xspace}

\usepackage{pgfplots}
 \pgfplotsset{compat=newest}
    \pgfplotsset{plot coordinates/math parser=false}
    \pgfplotsset{
    label style={anchor=near ticklabel},
    xlabel style={yshift=0.0em},
    ylabel style={yshift=-0.3em},
    tick label style={font=\footnotesize },
    label style={font=\footnotesize},
    legend style={font=\footnotesize},
    title style={font=\fontsize{7}}}

\usepackage{mathtools}
\mathtoolsset{showonlyrefs}
\usepackage{amssymb}
\usepackage{placeins}

\usepackage{multirow}
\usepackage{etoolbox}

\usepackage{balance} 
\usepackage{placeins} 

\usepackage{algorithm}
\usepackage[noend]{algpseudocode}

\usepackage{subfig}
\usepackage[normalem]{ulem}
\usepackage{multicol}

\usepackage[english]{babel}

\usepackage{amsthm}
\newtheorem{example}{Example}
\newtheorem{remark}{Remark}

\usepackage{booktabs}
\usepackage{ulem}

\usepackage[underline=true]{pgf-umlsd}
\usetikzlibrary{calc}

\usetikzlibrary{positioning}

\newcommand{\protocol}{rate-compatible protocol~}

\newcommand{\avgdi}{\bar{\a}_i}

\newcommand{\x}{\mathrm{x}}
\newcommand{\cntypes}{d_c}
\newcommand{\dntypes}{d_d}

\newcommand{\dnp}{\bm{p}}

\newcommand{\cnfraci}{f_i}

\newcommand{\matA}{\bm{D}}
\renewcommand{\a}{d}
\renewcommand{\S}{\mathcal{S}}
\newcommand{\D}{\Delta}
\renewcommand{\d}{\delta}
\newcommand{\SA}{\S_A}
\newcommand{\SB}{\S_B}
\newcommand{\univ}{\mathcal{U}}

\newcommand{\iblt}[1]{{\bm{c}}^{\{{#1\}}}}

\newcommand{\countfield}{\emph{count}\xspace}
\newcommand{\payload}{\emph{data}\xspace}

\newcommand{\hashlength}{\nu}
\newcommand{\keylength}{\kappa}

\newcommand{\hash}{H_{\bm{m},\bm{d}_j}}

\newcommand{\hashtype}{h_{\dnp}}

\newcommand{\rhoi}{\rho_{i}}
\newcommand{\rhoik}{\rho_{i,k}}
\newcommand{\lambdaij}{\lambda_{i,j}}

\newcommand{\hashx}{g}

\newcommand{\xor}{\text{XOR}\xspace}
\newcommand{\ispure}{\text{IsPure}\xspace}

\newcommand{\jvn} {g} 
\newcommand{\jv} {\bm{\jvn}} 
\newcommand{\jvi}[1]{\jvn_{#1}}

\newcommand{\msum}[1]{ m^{\star}}

\newcommand{\graph}{\mathcal{G}}
\newcommand{\cells}{\mathcal{C}}
\newcommand{\pairs}{\mathcal{Z}}
\newcommand{\edges}{\mathcal{E}}

\newcommand{\cell}{c}

\newcommand{\pair}{z}

\newcommand{\cellnode}{\mathtt{c}}
\newcommand{\pairnode}{\mathtt{z}}

\newcommand{\charpol}{\mathcal{X}}

\newcommand{\eff}{\eta}

\newcommand{\dest}{\hat{\d}}

\newcommand{\elsize}{\ell}
\newcommand{\itnumber}{\iota}

\newcommand{\app}{\gamma}

\newcommand{\pcn}[1]{\bar{w}_i^{(#1)}}
\newcommand{\pcni}[2]{\bar{w}_{#1}^{(#2)}}
\newcommand{\pvn}[1]{q_{i,j}^{(#1)}}
\newcommand{\pvnavg}[1]{\bar{q}_i^{(#1)}}

\newcommand{\w}{r}

\newtheorem{definition}{Definition}

\begin{document}

	\begin{acronym}
\acro{CPI}{characteristic polynomial interpolation}
\acro{LDPC}{low-density parity-check}
\acro{MET} {multi-edge-type}
\acro{IBLT}{invertible Bloom lookup table}
\acro{p.m.f.}{probability mass function}
\acro{IRSA}{irregular repetition coded slotted ALOHA}
\end{acronym}

\title{A rate-compatible solution to the set reconciliation problem}

\author{
	Francisco L\'azaro, Bal\'azs Matuz \\
		Institute of Communications and Navigation of DLR (German Aerospace Center),
		\\Wessling, Germany. Email: \{Francisco.LazaroBlasco, Balazs.Matuz\}@dlr.de
		\thanks{Copyright © 2023 IEEE. Personal use of this material is permitted.
			However, permission to use this material for any other purposes must be
			obtained from the IEEE by sending a request to pubs-permissions@ieee.org}
	}

\maketitle



\thispagestyle{empty} \pagestyle{empty}

	\begin{abstract}		

 	We consider a set reconciliation setting in which two parties hold similar sets that they would like to reconcile.
 	In particular, we focus on set reconciliation based on \acfp{IBLT}, a  probabilistic data structure inspired by Bloom filters. \ac{IBLT}-based set reconciliation schemes have the advantage of exhibiting low computational 
 complexity, however, the schemes available in the literature are known to be far from optimal in terms of communication complexity (overhead).
 	The inefficiency of \ac{IBLT}-based set reconciliation  can be attributed to two facts. First, it requires an estimate of the cardinality of the difference between the sets, which implies an increase in overhead.
 	Second, to cope with uncertainties in the estimation of the cardinality of the set difference,  \ac{IBLT} schemes in the literature oversize the data structures, thus further increasing the overhead.
 	In this work, we present a novel \ac{IBLT}-based set reconciliation protocol that does not require  estimating the cardinality of the set difference.
 	The proposed scheme relies on what we termed \acf{MET} \acp{IBLT}. 
 	The simulation results illustrate that the novel scheme outperforms state-of-the-art \ac{IBLT}-based approaches to set reconciliation in terms of communication cost, i.e., in terms of the number of bits to be exchanged.
	\end{abstract}

	\maketitle

	\section{Introduction}\label{sec:Intro}

\subsection{Motivation and Problem Description}\label{sec:setup}
Set reconciliation problems arise frequently in distributed systems 
with multiple copies of data across different nodes. Examples are redundant distributed databases, such as Amazon's Dynamo \cite{decandia2007dynamo} or Apache Cassandra \cite{cassandra}, in which copies of the same database are stored in multiple data centers (nodes). Whenever a new entry is added to or deleted from the database, the change is propagated to all the nodes in the system. This may, however, fail in some situations, for instance when some of the nodes are unavailable due to a hardware failure, or when communication is impaired by network congestion. The task of restoring the consistency among different copies of the database can be cast as a set reconciliation problem.
Set reconciliation protocols find also applications in distributed storage systems and remote file synchronization.  For the latter application, the simplest configuration consists of a setup with two copies of a file at different network locations. When one of the copies is updated, one would like to update the other copy, while transmitting as little data as possible to the remote network location. This problem can be converted to a set reconciliation problem by relying on tools from graph theory \cite{agarwal2006bandwidth}. 
Set reconciliation protocols can also be applied in gossip-spreading protocols \cite{MitzenmacherP18}. Recently, similar techniques have been proposed to address the problem of block propagation in a cryptocurrency network \cite{Ozisik2019}, as well as to synchronize the \emph{mempools}\footnote{A mempool is a database of pending Bitcoin transactions, which have not yet been written into the {Blockchain}.} of nodes in the Bitcoin network \cite{bovskov2022gensync}.%

In its simplest form, the set reconciliation problem can be described as follows. Two hosts $A$ and $B$ possess (possibly) different sets, $\SA$ and $\SB$, respectively. The goal is to determine the set difference $\D$ between  the two sets, i.e.,  the set of elements that are present in the union of $\SA$ and $\SB$,  but not in their intersection, $\D= (\SA \cup \SB) \setminus (\SA \cap \SB)$. To reconcile their sets, $A$ and $B$ exchange information  and perform certain computations. Thereby, both communication complexity (overhead) and computational (time) complexity shall be kept as small as possible. 
Intuitively, when the cardinality of the set difference $\d= |\D|$  is large, $A$ and $B$ may simply exchange their sets. However, this solution turns out to be highly sub-optimal when $\d$ is small. In this case,  more sophisticated algorithms can be applied to reduce communication complexity. 

A common approach to set reconciliation (and other similar problems) is the use of log files. The prerequisite here is the existence of  a so-called prior context between the parties involved in the protocol. In our setting, this translates to the assumption  that  sets $\SA$ and $\SB$ were synchronized (i.e., identical) at a previous point in time $\tau$. After that, both hosts keep track respectively of all the changes in their sets by using local log files. After $d$ units of time have elapsed, i.e., at time $\tau + d$, the two sets can be reconciled relying on the log files.
While being simple, the use of log files becomes inefficient when the network connectivity is bad, the storage medium is unreliable, or when changes to the sets $\SA$ and $\SB$ happen very frequently. Finally, these schemes do not scale well in multi-party settings (when one aims at synchronizing more than 2 sets), given that they require the existence of prior context between all pairs of nodes. Our focus instead will be on context-free set reconciliation.

\subsection{Related Works}

Fundamental limits for one-way set reconciliation, i.e., when one single communication message is allowed, were studied in \cite{Karpovsky2003}. Let the elements in $\SA$ and $\SB$ be $\elsize$ bit vectors, i.e., $\SA, \SB\subset \{0,1\}^\elsize$, 
and let an upper bound $t$ to the cardinality (or size) of the set difference $\d=|\D|$ be known, i.e., $\d \leq t$. Then, exact set reconciliation requires transmitting $\sim t \elsize$ bits. In contrast, when no upper bound is known, exact set reconciliation requires transmitting the whole set. 
Given an upper bound $t$ to the set difference $\d$,  set reconciliation can be carried out relying on  a binary code of length $2^\elsize$ which can correct  $t$ errors. This solution is shown to have an optimal communication complexity when using a perfect code \cite{Karpovsky2003}. However, it requires working with a code of length $2^\elsize$ which is impractical for the values of $\elsize$ required by most applications. 

In \cite{Minsky2003} a scheme based on \acf{CPI} was proposed which achieves nearly optimal communication complexity. In particular, when an upper bound $t$ to the size of the set difference $\d$ is known, exact set reconciliation is possible with communication complexity  $\sim t\elsize$ and time complexity $\mathcal{O}(t^3)$. 
When an upper bound to $\d$ is not known, \cite{Minsky2003} proposes an approximate\footnote{By approximate it is meant that the scheme may fail, so that the parties wrongly believe that their sets are reconciled, but they are actually not. In practice, the scheme can be designed to make this failure probability arbitrarily low, at the cost of an increase in the communication cost.} set reconciliation scheme. In this case, the scheme requires an additional feedback message and is thus no longer a one-way reconciliation protocol. Instead, we have a reconciliation scheme that requires one round of communication. This approximate scheme has communication complexity $\sim \d\elsize$ and time complexity $\mathcal{O}(\d^4)$.

In \cite{eppstein:2011}  a low-complexity approximate set reconciliation scheme based on  \acfp{IBLT} was proposed, which also requires one communication round.
In brief, an \ac{IBLT} \cite{goodrich:2011} is a probabilistic data structure that  supports the insertion and deletion of set elements, as well as listing of all the elements contained in the \ac{IBLT}, provided that the number of elements in the \ac{IBLT} is not too large (a formalization of these concepts follows in Section~\ref{sec:IBLT}). From a data structure point of view, an \ac{IBLT} can be broken down into cells, which have size $\sim \elsize +\log{\elsize}$. 
When an upper bound to the  size of the set difference is known, the communication complexity of \ac{IBLT}-based set reconciliation is $\sim c \, t (\elsize+\log \elsize)$, where $c\approx 1.22$, and the  time complexity is $\mathcal{O}(t (\elsize + \log \elsize ))$ \cite{goodrich:2011}. 

\ac{IBLT}-based set reconciliation requires inverting an \ac{IBLT}, an operation that is identical to erasure decoding \cite{Eli54}. As such, different \ac{IBLT} designs have been proposed in the literature which are akin to different erasure code designs. The original construction in \cite{eppstein:2011} relies on \emph{regular} \acfp{IBLT} in which each set element is mapped into the same number of  \ac{IBLT} cells. By moving to \emph{irregular}  \acp{IBLT} it is possible to improve the constant $c$. In particular, it was first shown in \cite{rink2013mixed} that $c=1.09$ can be achieved by relying on irregular \acp{IBLT} with two different degrees. In \cite{lazaro21:iblt} a density evolution analysis of arbitrary irregular \acp{IBLT} was presented which allows  computing $c$ and thus optimizing the irregular \ac{IBLT}. Furthermore, relying on results for random access protocols \cite{narayanan2012iterative}, it was argued that \acp{IBLT} with constants $c$ arbitrarily close to one can be found.  

A shortcoming of \ac{IBLT}-based set reconciliation \cite{eppstein:2011} is the fact that it requires knowledge of an upper bound $t$ to the size of the set difference $\d$. 
In most cases, rather than an upper bound to $\d$, an (imperfect) estimate of $\d$ is available, denoted by $\dest$. Obtaining this estimate usually requires  an additional communication round  between the hosts,  which represents an additional overhead \cite{eppstein:2011}. In practice, the  estimated set difference size $\dest$ can substantially differ from the actual one $\d$. Intuitively, an over-estimation ($\dest > \d$) results in an oversizing of the \ac{IBLT} and thus in an unnecessarily high communication cost, whereas an under-estimation ($\dest < \d$) results in  a high probability of failure of the set reconciliation algorithm. To reduce the probability of a failure, \cite{eppstein:2011} considers an oversizing of the \ac{IBLT}. Additionally, a larger \ac{IBLT} can be sent upon a failure, attempting  set reconciliation using only this second \ac{IBLT}. This approach is inefficient since it discards the first exchanged \ac{IBLT}. This inefficiency was identified in \cite{Ozisik2019}, where it was proposed to use the two \acp{IBLT} to reconcile the sets, reducing the communication cost.

Both \ac{CPI} and \ac{IBLT}-based set reconciliation assume that elements in $\SA$ and $\SB$ have all equal length, i.e., that they are length-$\ell$ bit vectors. However, in some applications, the set elements might have variable lengths. For example, the set elements might be encoded using strings of length ranging between $\ell_{\text{min}}$ and $\ell_{\text{max}}$ bits. In this case, a possible workaround is padding all elements to the maximum length $\ell_{\text{max}}$. This yields an increase in the overhead (communication cost). Alternatively, one may rely on other techniques for approximate set reconciliation protocols that can inherently deal with variable-length data. Multiple such approaches exist based on so-called approximate set membership data structures. In \cite{Skjegstad2011} a low-complexity  multi-round protocol based on the exchange of Bloom filters was proposed. Protocols based on the exchange of counting Bloom filters and Cuckoo filters were proposed in \cite{Guo2013} and \cite{Luo2019}, respectively. 
In \cite{kruber2020approximate} approximate set reconciliation protocols based on the exchange of hash lists, Bloom filters, and Merkle trees were proposed and compared.
For a comparison of the different set reconciliation protocols in different practical settings, we refer the reader to \cite{bovskov2022gensync}.

\vspace{-6mm}
\subsection{Contribution}

\vspace{-2mm}
In this work, we introduce a novel approach to \ac{IBLT}-based set reconciliation. In contrast to previous works, our approach does not require an upper bound to (or an estimate of) the size of the set difference $\d$. Our work borrows ideas from fountain codes \cite{byers02:fountain}. In particular, we let the transmitter (say host $A$) send the cells of the \ac{IBLT} one by one. After receiving some cells, the receiver (host $B$) can attempt to reconcile the sets. If set reconciliation fails, the receiver simply waits for more cells and re-attempts set reconciliation. When set reconciliation is successful, the receiver sends an acknowledgment to the transmitter so that it stops sending cells. 
This approach spares the overhead necessary for set difference estimation which is considerable, especially when the set difference is small \cite{eppstein:2011}. %
To fully exploit the advantages of our approach, we introduce a novel \ac{IBLT} structure which we term \acf{MET} \ac{IBLT}. This novel data structure borrows ideas from rate-compatible codes, in particular from \ac{MET} \ac{LDPC} codes \cite{richardson2002multi}.
The main advantage of the adoption of a \ac{MET} structure is that it allows increasing the number of \ac{IBLT} cells on demand while still being able to list all the elements with high probability. This is akin to the feature of rate-compatible codes of operating close to capacity at multiple code rates.

The remaining part of the paper is structured as follows. In Section~\ref{sec:IBLT} we introduce the basics of \ac{MET} \acp{IBLT}. In Section~\ref{sec:analysis} we
develop an analysis to assess the communication efficiency of \ac{MET} \acp{IBLT}. We sketch a flexible set reconciliation protocol using \ac{MET} \acp{IBLT} in Section~\ref{sec:setrec_prot} followed by a design example in Section~\ref{sec:design} and complementary simulation results. Section~\ref{sec:conclusions} provides a discussion of the contribution.

\section{\acl{MET} \aclp{IBLT}}\label{sec:IBLT}

\subsection{Definitions}
We first introduce a few definitions.
\begin{definition}[Key-value pair]
A key-value pair $\pair$ is a data structure composed of two fields, a key $x$  of length $\hashlength$ bits and a value $y$ of length {$\keylength$ bits, where typically $\keylength\gg \hashlength$}. The key $x$ is obtained as a function of $y$, $x=\hashx (y)$ where the mapping is many to one.\footnote{This means that key collisions are in principle possible. However, by choosing $\hashlength$ large enough, one can make the key collision probability small enough for practical applications.} 
\end{definition}
The term key-value pair has its origin in the field of database systems.
In particular, in a so-called key-value database every database entry is a key-value pair.
In this context,  the key can be thought of as a short (unique) identifier of an element in the database, whereas the value is the actual data which can be orders of magnitude larger than the key.  In our case, every key-value pair will represent a set element, and we will assume that the key associated with a set element is obtained as a hash function of the set element's value.

\begin{definition}[Cell]
A cell $\cell$ is a data structure containing two different fields, \payload and  \countfield where:
\begin{itemize}
    \item \payload=(\payload.x, \payload.y) is a bit string of length $\hashlength+\keylength$. 
    The bit strings \payload.x and \payload.y contain, respectively, the  binary \xor of the keys and values that have been mapped to the cell.
    \item \countfield is an integer. It contains the number of key-value pairs that have been mapped to this cell (see Section~\ref{sec:MET_description}).
\end{itemize}
\end{definition}
\begin{definition}[IBLT]\label{def:IBLT}
An \acf{IBLT}  is a data structure that is used to represent a set of key-value pairs. Formally, it consists of an array of $m$ cells $\bm{c}=\begin{bmatrix} c_1 &  c_2 & \dots & c_m \end{bmatrix}$. The choice of the hash functions (Section~\ref{sec:MET_description}) determines how key-value pairs are mapped into the \ac{IBLT} cells.
\end{definition}
{ \acp{IBLT} support multiple operations, such as the insertion and deletion of key-value pairs, as well as the listing of all inserted key-value pairs (with high probability).}
{For simplicity, and for being consistent with the literature, we will sometimes abuse $\bm{c}$ to refer to the \ac{IBLT} itself, rather than only to its $m$ cells. Furthermore, we will sometimes say that we `transmit an \ac{IBLT}' when we actually mean that we transmit the array of cells $\bm{c}$.}

\begin{definition}[MET IBLT]\label{def:METIBLT}
{A \acf{MET} \ac{IBLT} is a generalization of an \ac{IBLT} that assigns different labels or types to key-value pairs and cells. }
\end{definition}
The notion of types allows putting specific constraints on the mapping of key-value pairs to cells. 
From a data structure point of view, introducing a \ac{MET} structure allows addition of new groups of cells  on demand. Furthermore, by judiciously choosing the hash functions used to map key-value pairs into cells we can improve the performance of the recovery operation of our data structure (see Section~\ref{sec:analysis}). 
In Section~\ref{sec:design} we will highlight the advantages of \ac{MET} \acp{IBLT} used for set reconciliation compared to unstructured \acp{IBLT} which can be seen as a \ac{MET} \ac{IBLT} with one single key-value pair type and one single cell type.

\subsection{MET IBLT Description} \label{sec:MET_description}
We assign to each cell and key-value pair a label referred to as type which will play a role {in how the mapping of key-value pairs into cells is done}. Let $\cntypes$ be the number of different cell types, $m_i$, $i \in \{1,2,\ldots,\cntypes\}$ be the number of cells of type $i$, and $m=\sum_i m_i$ be the length (number of cells) of the \ac{MET} \ac{IBLT}. For convenience, we assume that the cells are ordered according to their type, so that cells $\cell_1, \cell_2, \dots, \cell_{m_1}$ are of type 1, cells $\cell_{m_1+1}, \cell_{m_1+2}, \dots, \cell_{m_1+m_2}$  are of type 2, and so forth. 

Further, let the total number of different key-value pair types be $\dntypes$. The type of a key-value pair $\pair$ is determined by applying a hash function $\hashtype(\pair.x)$ to the key $\pair.x$, where $\hashtype(\pair.x)=j$ maps an input  $\pair.x \in \{0,1\}^{\hashlength}$ to an output $j \in \{1,2,\dots,\dntypes\}$. This hash function 
is parameterized by the probability vector ${\dnp=\begin{bmatrix} p_1 & p_2& \dots & p_{\dntypes}\end{bmatrix}}$. 
The vectors $\begin{bmatrix} 1 &2 & \ldots &\dntypes\end{bmatrix}$ and $\dnp$ describe a \ac{p.m.f.}, the so-called \emph{type distribution}. 
 In particular, under the assumption that the input $\pair.x$ is uniformly distributed, the output of $\hashtype(\pair.x)$ samples the type distribution.

The mapping of a key-value pair of type $j$ to cells of type $i$ is achieved through the hash function $\hash(\cdot)$. It is parameterized by $\bm{m}= \begin{bmatrix} m_1 & m_2 & \dots & m_{\cntypes}\end{bmatrix}$ and $\bm{\a}_j=\begin{bmatrix}\a_{1,j} & \a_{2,j}& \dots & \a_{\cntypes,j}\end{bmatrix}$ and is used to sample at random $\a_{i,j}$ cells of type $i$ into which the key-value pair of type $j$ is mapped. The function outputs a vector $\jv=\hash(\pair.x)$ of length $\sum_{i=1}^{\cntypes} \a_{i,j}$, which contains $\a_{i,j}$ indices of cells of type $i$ (where the indices are between $1+\sum_{k=1}^{i-1} m_k$ and $\sum_{k=1}^{i} m_k$).
\begin{example}[Hash function $\hash$] \label{example:hash}
    Assume $\bm{m}=\begin{bmatrix}2 & 3& 5 \end{bmatrix}$ and $\bm{\a}_j=\begin{bmatrix}1 & 1& 2\end{bmatrix}$. 
    We have $m_1=2$ cells of type 1, $m_2=3$ cells of type 2 and  $m_3=5$ cells of type 3. 
    Hence, indices $1$ and $2$ are associated with cells of type 1. 
    Indices $3$, $4$, and $5$ are associated with cells of type 2, and indices $6,7,8,9$, and $10$ are associated with cells of type $3$. The hash function returns {4 indices in total, }one index associated with a cell of type 1, one index associated with a cell of type 2, and two indices associated with cells of type 3.
    For a first key $x_1$, the hash function $\hash(x_1)$ may return  $\hash(x_1)=\begin{bmatrix} 1 & 4& 7& 9 \end{bmatrix}$. For a second key $x_2$, the hash function $\hash(x_2)$ may return  $\hash(x_2)=\begin{bmatrix} 2 & 3& 6& 7\end{bmatrix}$.
\end{example}

\ac{MET} \acp{IBLT} support different operations that are discussed in more detail in the next sections:
\begin{itemize}
	\item Initialize$()$. This operation sets the different fields  of all the cells in the \ac{IBLT} to zero (see Algorithm~\ref{alg:init}).
	\item Insert$(\pair)$. The insertion operation \emph{adds} the key-value pair $\pair$ to the \ac{IBLT} following certain mapping rules (see Algorithm~\ref{alg:insert}).
	\item Delete$(\pair)$. The deletion operation \emph{removes} the key-value pair $\pair$ from the \ac{IBLT} (see Algorithm~\ref{alg:deletion}).
	\item Recover$()$. This operation aims at listing all key-value pairs stored in the \ac{IBLT}. If this operation provides the full list of key-value pairs in the \ac{IBLT}, we say it succeeds, else it fails (see Algorithm~\ref{alg:list}).
\end{itemize}

\begin{figure}

\begin{minipage}{0.475\textwidth}
\begin{algorithm}[H]
	\small
	\caption{Initialization}\label{alg:init}
	\begin{algorithmic}
		\vspace{-1.5mm}
		\Procedure{Initialize$()$}{}
		\For {i = $1,2,\dots, m$} 
		\State $\cell_{ { i } }.\countfield =0$
		\State $\cell_{ { i } }.\payload =\bm{0}$
		
		\EndFor
		\EndProcedure
	\end{algorithmic}
\end{algorithm}
\end{minipage}\hfill
\begin{minipage}{0.475\textwidth}
\begin{algorithm}[H]
			\small
			\caption{Insertion}\label{alg:insert}
			\begin{algorithmic}[H]
				\vspace{-1.5mm}
				\Procedure{Insert$(\pair)$}{}
				\State $j \gets \hashtype(\pair.x)$
				\State $ \jv \gets \hash(\pair.x)$
		        \For {k = $1,2,\dots, \text{length}(\jv)$} 
		            \State $\cell_{ \jvi{ k } }.\countfield = \cell_{ \jvi{ k } }.\countfield +1$
		            \State $\cell_{ \jvi{ k } }.\payload =\xor \left( \cell_{ \jvi{ k } }. \payload, \,  \pair \right)$
		            \EndFor
				\EndProcedure
			\end{algorithmic}
		\end{algorithm}
\end{minipage}
\begin{minipage}{0.475\textwidth}
	\begin{algorithm}[H]
			\small
			\caption{Deletion}\label{alg:deletion}
			\begin{algorithmic}
				\vspace{-1.5mm}
				\Procedure{Delete$(\pair)$}{}
			    \State $j \gets \hashtype(\pair.x)$
				\State $ \jv \gets \hash(\pair.x)$
		        \For {k = $1,2,\dots, \text{length}(\jv)$} 
		            \State $\cell_{ \jvi{ k } }.\countfield = \cell_{ \jvi{ k } }.\countfield -1$
		            \State $\cell_{ \jvi{ k } }.\payload =\xor \left( \cell_{ \jvi{ k } }. \payload, \,  \pair \right)$
		            \EndFor
			\EndProcedure
			\end{algorithmic}
		\end{algorithm}
\end{minipage}\hfill
\begin{minipage}{0.475\textwidth}
\begin{algorithm}[H]
	\small
	\caption{Recovery}\label{alg:list}
	\begin{algorithmic}
		\vspace{-1.5mm}
		\Procedure{Recover$()$}{}
		
		\While {$\exists i \in [1, m] | \cell_{i}.\countfield = 1  $} 
		\State add $\pair=\cell_{i}.\payload$ to the output list
		\State call Delete $(\pair)$
		\EndWhile
		\EndProcedure
	\end{algorithmic}
\end{algorithm}
\end{minipage}
\end{figure}

\vspace{-3mm}
\subsection{Insertion of  a set $\S$ into a \acs{MET} \acs{IBLT}}
We consider now the insertion of a set $\S=\{\pair_1, \pair_2, \dots, \pair_n\}$ of $n$ elements (key-value pairs) into a \ac{MET}  \ac{IBLT} that has been previously initialized to zero by Algorithm~\ref{alg:init}. The elements of $\S$ are successively inserted  into the \ac{IBLT} as described by Algorithm~\ref{alg:insert}: first, every  element $\pair=(x,y)$ is assigned a data node type $j= \hashtype(\pair.x)$. Next, the hash function $\hash(\pair.x)$ is used to obtain a vector $\jv$ that contains the indices of the cells into which $\pair$ is to be mapped, i.e., those cells whose  \payload field is \xor-ed with $\pair$. The respective $\countfield$ fields are increased by one.

\subsection{Recovery of the set $\S$}

The recovery operation is a low-complexity algorithm that aims at inverting the \ac{IBLT}, i.e., its goal is to recover all the key-value pairs that have been inserted into the \ac{IBLT}. Algorithm~\ref{alg:list} describes the simplest form of the recovery operation.  
This algorithm consists of seeking cells with \countfield field equal to one. Such cells contain a single element $\pair$ of $\S$, which can be directly obtained from their \payload field. Thus, one can extract the element $\pair$ and append it to the output list.
After that, one deletes $\pair$ from the \ac{IBLT} by calling Delete$(\pair)$, which effectively removes $\pair$ from all the cells in which it had been mapped and reduces the count fields by one. 
This process is repeated until no more cells with \countfield equal to one can be found. Recovery succeeds if at the end all $m$ cells of the \ac{IBLT} have  \countfield field  equal to zero. 
If some cells have a non-zero count,  recovery fails. 

As stated in \cite{mitzenmacher:2012}, the recovery operation is an instance of \textit{peeling decoding} \cite{luby1998analysis}. 
Peeling decoding is best understood considering a graph representation of an \ac{IBLT}. In particular, we may represent an \ac{IBLT} with $m$ cells in which a set $\S$ of $n$ elements (key-value pairs) have been inserted as a bipartite (or Tanner) graph $\graph=(\pairs \cup \cells, \edges)$ composed of a set of $n$ data nodes $\pairs$, a set of $m$ cell nodes $\cells$ and a set of edges $\edges$. 
As the names indicate, data nodes represent key-value pairs and cell nodes represent cells of the \ac{IBLT}. 
A data node $\pairnode_i \in \pairs$ and a cell node $\cellnode_h \in \cells$ are connected by an edge if and only if  $\pair_i=(x_i,y_i)$ is mapped into cell $\cell_h$. 
A  data node $\pairnode$ and a cell node $\cellnode$ are said to be neighbors if they are connected by an edge. We use the shorthand $\mathcal{N}(\cellnode)$ or $\mathcal{N}(\pairnode)$ to denote the set of all neighbors of a cell node $\cellnode$ or a data node $\pairnode$. The degree of a node is given by the number of edges connected to the node. Thus, the degree of a cell node equals the $\countfield$ field of the cell it represents.

\begin{example}[Graph representation]
Figure~\ref{fig:graph_repr} shows the graph representation of an \ac{IBLT} with $m=4$ cells in which $n=3$ pairs have been inserted. 
Key-value pair $\pair_1$ is mapped into cells $\cell_1$ and $\cell_4$, key-value pair $\pair_2$ is mapped into cells $\cell_1$ and $\cell_2$ and key-value pair $\pair_3$ is mapped into cells $\cell_2$ and $\cell_4$. 
Thus, we have $\countfield= 2$ at cells $\cell_1, \cell_2$, and $\cell_4$, and  $\countfield= 0$ in $\cell_3$.

In the graph representation of the \ac{IBLT}, each key-value pair $\pair$ is represented by a data node $\pairnode$, shown as a circle in the figure. 
Each cell $\cell$ is represented by a cell node $\cellnode$, shown as a square.  
An edge connects a data node $\pairnode_i$ to a cell node $\cellnode_h$ if and only if  $\pair_i$ is mapped into cell $\cell_h$. For example, $\pairnode_1$ is connected to $\cellnode_1$ and $\cellnode_4$.
We can also observe that the degree of a cell node $\cellnode$ (the number of edges attached to it) equals the $\countfield$ field of the cell it represents. Thus, cell nodes $\cellnode_1, \cellnode_2$, and $\cellnode_4$ have degree 2, whereas $\cellnode_3$ has degree zero.

\begin{figure}[t!]
    \centering
    \includegraphics[width=0.9\columnwidth]{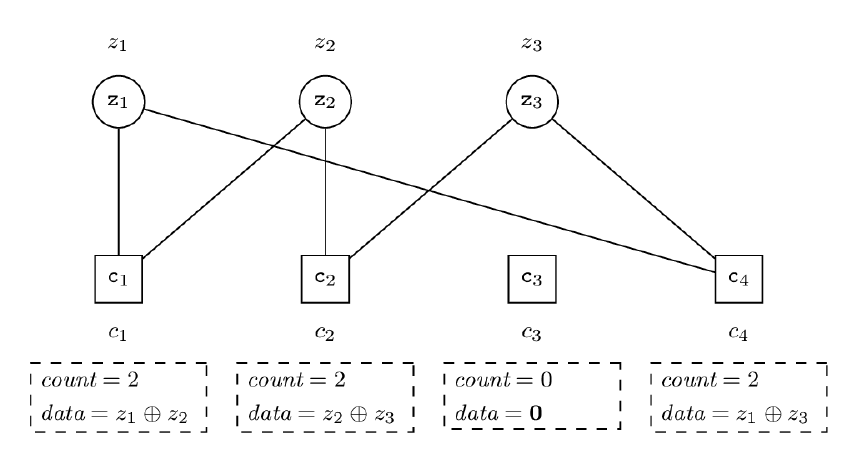}
    \caption{Graph representation of an \ac{IBLT}. The data nodes $\pairnode$ are represented by circles and the cell nodes $\cellnode$ by squares. An edge connects a data node $\pairnode_i$ to a cell node $\cellnode_h$ if and only if  $\pair_i$ is mapped into cell $\cell_h$. The degree of a cell node $\cellnode$ equals the $\countfield$ field of the cell it represents. The operator  $\oplus$ denotes a bit-wise XOR.}
    \label{fig:graph_repr}
\end{figure}
\end{example}

Recovery of $\S$ can be represented as a peeling process on a bipartite graph. In particular, whenever a cell node $\cellnode$ of degree one is present, its only neighbor $\mathcal{N}(\cellnode)=\pairnode$ is determined. The key-value pair $\pair$ which is represented by the data node $\pairnode$ is added to the output list. Next, the retrieved key-value pair is deleted from  the \ac{IBLT}. This translates into the removal of all edges attached to its associated data node. 
This process is repeated until no more cell nodes of degree one are present. At this stage, if all cell nodes are of degree zero, recovery succeeded and all key-value pairs are present in the output list. Otherwise, if some cell nodes of a degree larger than zero are present, recovery fails, and the output list will not contain all key-value pairs. Note that  the graph structure is unknown to the decoder, and is revealed successively during the decoding process (see Examples~\ref{example:peeling} and \ref{example:peelingII}).  

\begin{figure}[t!]	
	\subfloat[$\itnumber=0$]{
		\includegraphics[width=0.5\columnwidth]{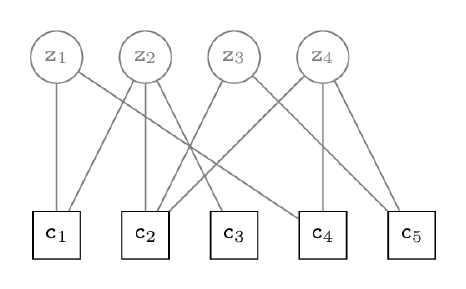}
		
		\label{fig_l_0}
	}
	\subfloat[$\itnumber=1$]{
		\includegraphics[width=0.5\columnwidth]{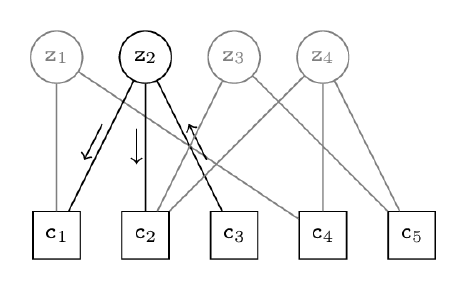}
		\label{fig_l_1}
	}

	\subfloat[$\itnumber=2$]{
		\includegraphics[width=0.5\columnwidth]{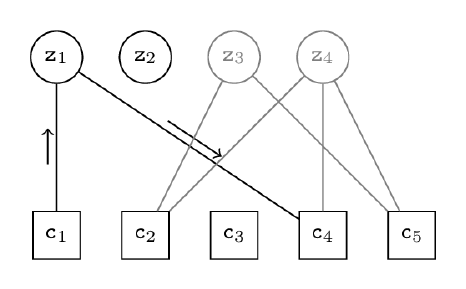}
		\label{fig_l_2}
	}    
	\subfloat[$\itnumber=3$]{
		\includegraphics[width=0.5\columnwidth]{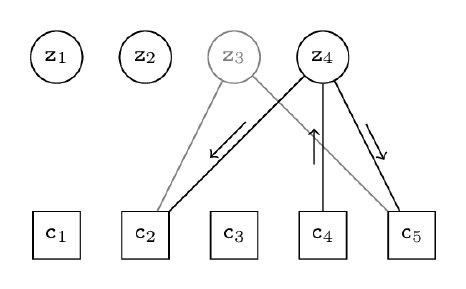}
		\label{fig_l_3}
	}	
	\begin{center}	
	\subfloat[$\itnumber=4$]{
			\includegraphics[width=0.5\columnwidth]{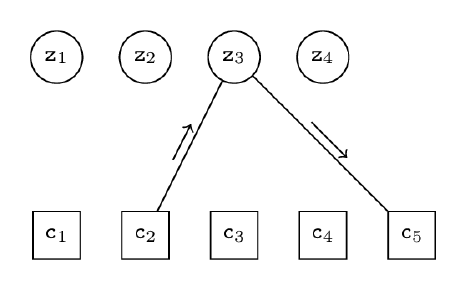}			
			\label{fig_l_4}
	}
	\end{center}

	\caption{Peeling process on the graph representation of an \ac{IBLT} with $n=4$ key-value pairs and $m=5$ cells. 
    The index $\itnumber$ is used to denote the different steps of the recovery process. The data nodes and edges shown in gray are those which, at the current iteration, are still unknown to the decoder. The data nodes shown in black have already been recovered by the decoder at the end of the current iteration $\itnumber$. The arrows parallel to the edges indicate the traversal direction of the peeling process.
    }
	\label{fig:graph_peeling}
\end{figure}

\begin{example}[Peeling decoding I] \label{example:peeling}
The different steps of the peeling process are shown in Figure~\ref{fig:graph_peeling}. Figure~\ref{fig_l_0} shows the bipartite graph representation of an \ac{IBLT} before the peeling process starts. We observe that  the \ac{IBLT} has  $m=5$ cells and it stores  $n=4$  key-value pairs. 
However, at this stage, the depicted bipartite graph is \textit{unknown} to the decoder since it does not have any knowledge about $\S$. The decoder is only aware of the $m$ cell nodes. For this reason, the data nodes, as well as the edges, are shown in gray. The graph structure will be revealed successively as the recovery operation progresses, and it will only be completely known if decoding succeeds. Otherwise, a part of the graph will remain hidden.
We can  see that cell node $\cellnode_3$ has degree $1$, and thus its associated \ac{IBLT} cell $\cell_3$ has count $1$.  The recovery operation retrieves the only key-value pair that has been mapped to cell $\cell_3$, i.e., data node $\pair_2$, which is added to the output list of the recovery operation. Afterwards, $\pair_2$ is deleted from the \ac{IBLT}. In the graph representation, this translates to revealing the only neighbor of cell node $\cellnode_3$, data node $\pairnode_2$ (now shown in black), and deleting all edges attached to it, as shown in Figure~\ref{fig_l_1}. As a consequence, the degree of $\cellnode_1$ becomes one. 
In the next step, as shown in Figure~\ref{fig_l_2}, the only neighbor of $\cellnode_1$, $\pairnode_1$, is revealed and all edges attached to it are removed. This reduces the degree from  $\cellnode_4$ from $2$ to $1$.
Then, data node $\pairnode_4$ is revealed since it is the only neighbor of $\cellnode_4$. After all the edges attached to $\pairnode_4$ are removed, as shown in Figure~\ref{fig_l_3},  we have two cell nodes of degree $1$, namely $\cellnode_2$ and $\cellnode_5$, both of which have as only neighbor $\pairnode_3$.
In the last step shown in Figure~\ref{fig_l_4}, $\pairnode_3$ is revealed as the only neighbor of  $\cellnode_2$. Finally, all edges attached to $\pairnode_3$ are erased from the graph. In this example, the recovery operation succeeded, and set $\S$ was completely recovered.
\end{example}

\begin{example}[Peeling decoding II] \label{example:peelingII}
Consider now the application of peeling decoding to the graph in Figure~\ref{fig:graph_repr}. In contrast to  Example~\ref{example:peeling}, peeling decoding now fails. It cannot even get started since there are no cells with count $1$, i.e., there are no cell nodes of degree $1$.
\end{example}

	\section{Analysis}\label{sec:analysis}

We consider the case where the number of cells $m$ grows large and are interested in the probability that a random key-value pair cannot be recovered given that $n\leq m$ key-value pairs were previously stored in the \ac{MET} \ac{IBLT}. Our analysis is based on \ac{MET} density evolution \cite{richardson2002multi} that provides a general framework to obtain the asymptotic performance of \ac{LDPC} code ensembles. In the asymptotic regime, \ac{MET} density evolution allows to determine the \emph{average} probability of a key-value pair being present in the output list of the recovery operation, where the average is taken over an ensemble of bipartite graphs. 
Even while fixing the \ac{IBLT} parameters and hash functions, different graphs are obtained by inserting different sets $\S$ into the \ac{IBLT}. Therefore, we are interested in the average probability of recovery failure.

\subsection{MET IBLT description}
We associate to each cell node a node type $i \in \{1,2,\ldots,\cntypes\}$. Cell nodes of the same type may have \textit{different degrees}. We introduce the edge-oriented degree distribution polynomial for cell nodes of type $i$
\begin{align} \label{eq:rhoi}
\rhoi(\x)&=\sum_{k=1} \rhoik \x^{k-1}   
\end{align}
where the coefficients $\rhoik$ denote the number of edges emanating from cell nodes of type $i$ and degree $k$ divided by all edges emanating from cell nodes of type $i$. As we will shortly see, $\rhoi(\x)$ is induced by the node degree and type distributions of the data nodes.

Likewise, each data node is associated with a data node type  $j \in \{1,2,\ldots,\dntypes\}$. We use a $\cntypes \times \dntypes$  degree matrix $\matA=[\a_{i,j}]$ to  describe the connectivity between the different types of data and cell nodes, where $\a_{i,j}$ corresponds to the number of edges connecting a type-$j$ data node to a type-$i$ cell node (recall Section~\ref{sec:IBLT}). 
Considering only edges connected to cell nodes of type $i$ {(i.e, row $i$ of $\matA$)}, we define the expected data node degree $\avgdi$ from the perspective of a cell node of type $i$ as
\begin{align}\label{eq:analysis:avg_degree}
    \avgdi  & = \sum_{j=1}^{\dntypes} p_j \a_{i,j}. 
\end{align}
Further, considering still row $i$ of $\matA$ only, we denote by  $\lambdaij$ the fraction of edges connected to data nodes of type $j$, 
  $ {\lambda}_{i,j} = \frac{ p_j \a_{i,j}}{\avgdi}$.

\begin{figure}[t]
    \centering
    \includegraphics*[width=\columnwidth]{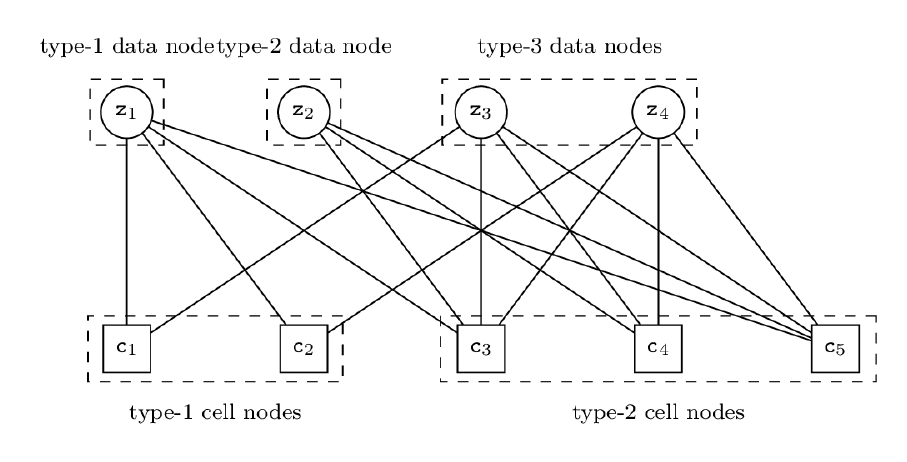}
    \caption{Bipartite graph example of a \ac{MET} \ac{IBLT} with $n=4$ data nodes 
    and $m=5$ cell nodes of different types.}
    \label{fig:MET_IBLT}
\end{figure}

Hence, ${\bm \lambda_i}=[{\lambda}_{i,j}],$ $j \in \{1,2,\ldots,\dntypes\}$ gives an edge-oriented data node degree distribution considering only edges connected to type $i$ cell nodes (and ignoring all other edges). 
\begin{example}\label{ex:MET1}
Given the degree matrix $\matA=\begin{bmatrix}    2    & 0 &    1\\
     2  &   3   &  3\end{bmatrix}$ and the probability vector $\bm p=\begin{bmatrix}    0.25    & 0.25 &    0.5\end{bmatrix}$. We find that there are three different data node types and two different cell node types. With probability $1/2$ a data node is of type three while with probability $1/4$ it is of type one or two, respectively. The average data node degrees from the cell node perspective are $\bar{\a}_1= 1.00$ and $\bar{\a}_2=2.75$. Further, the data node edge-oriented degree distributions from the cell node perspective are ${\bm \lambda}_1=\begin{bmatrix} 0.50   &      0   & 0.50\end{bmatrix}$ and ${\bm \lambda}_2=\begin{bmatrix}  0.18  &  0.27 &   0.55\end{bmatrix}$.
\end{example}
\begin{example}\label{ex:MET4}
	Assume a \ac{MET} \ac{IBLT} with the parameters from Example~\ref{ex:MET1}. In addition, let $n=4$, $m=5$, and $m_1=2$, $m_2=3$. Figure~\ref{fig:MET_IBLT} exemplifies a bipartite graph with such parameters.\footnote{The connections are data dependent and are chosen at will in this example, respecting the type constraints.}
\end{example}

\subsection{Load threshold computation}
\begin{definition}[Load]
Let the  {load} $\eff=n/m$, be the ratio between the number of key-value pairs and cells used to store those, i.e., the length of the \ac{IBLT}.
\end{definition}

Let us consider the regime for which $n$ and $m$ go to infinity. We are interested in the maximum load, the so-called load threshold $\eff^{\star}$ of a \ac{MET} \ac{IBLT} ensemble, such that recovery is successful {with high probability}. 
In the context of \ac{LDPC} codes, the threshold of code ensembles with peeling decoding is analyzed via \textit{density evolution} \cite{luby1998analysis,richardson2001design} which restates the peeling decoder as an \textit{equivalent} iterative message passing algorithm where nodes pass messages along the edges to their neighbors. The messages exchanged by the nodes can be either an \textit{erasure}, i.e., we do not know the corresponding key-value pair yet, or a \textit{non-erasure}, meaning that key-value pair is known.

Let us denote by $\pcn{\itnumber}$ the (average) probability that the message sent from a cell node of type $i$ over an edge at the $\itnumber$-th iteration is an erasure  and by $\pvnavg{\itnumber}$ the (average) probability that the message sent from a data node to a cell node of type $j$ over an edge at the $\itnumber$-th iteration is an erasure. 
\ac{MET} density evolution iteratively computes $\pcn{\itnumber}$ and $\pvnavg{\itnumber}$.

As a first step in our analysis, we need  the edge perspective degree distribution for cell nodes of type $i$ in the limit when $m\rightarrow \infty$. As shown in \cite{Liva2011}, it is
\begin{align} \label{eq:rho_partial}
     {\rho}_i(\x)=e^{- \frac{\eta}  {\cnfraci} \,  \avgdi \, (1-\x) }
\end{align}
where $\cnfraci\triangleq m_i/m$ is the fraction  of cell nodes of type $i$ and $\eta / {\cnfraci}=n/m_i$ is the load for the $m_i$ cell nodes of type $i$.
The message sent by a cell node of type $i$ will only be a non-erasure if all other incoming messages are  non-erasures, i.e.,
\begin{align}
   \pcn{\itnumber} & ={1- {\rho}_i\left(1- \pvnavg{\itnumber}\right)} 
   =1-e^{ - \frac{\eta}{\cnfraci} \avgdi  \pvnavg{\itnumber}} .
   \label{eq:de_cn_met}
\end{align}
In other words, if all but one key-value pairs mapped into a cell are known, also the remaining key-value pair can be determined (non-erasure message).
Let $\pvn{\itnumber}$ be the erasure probability for messages sent from a type-$j$ data node to a type-$i$ cell node at the $\itnumber$-th iteration. The message sent by a data node will only be an erasure if all other incoming messages are erasures, i.e.,
\begin{align}\label{eq:de_vn_met}
    \pvn{\itnumber} &= \prod_{k=1}^{\cntypes} \left( {\pcni{k}{\itnumber-1}}\right)^{b_{k,j}}
\end{align}
where $b_{k,j}=\a_{k,j}$ if $k \neq i$ else $b_{k,j}=\max(0,\a_{k,j}-1)$. In other words, if a key-value pair has been recovered from any cell, its value can be subtracted from all other cells where it has been previously mapped into. The average erasure probability  at the input of a cell node of type $i$ at the $\itnumber$-th iteration $\pvnavg{\itnumber}$ is given by
\begin{align} \label{eq:de_vn_avg_met}
    \pvnavg{\itnumber} &= \sum_{j=1}^{\dntypes} {\lambda}_{i,j}   \pvn{\itnumber}.
\end{align}

By initially setting $\pvnavg{0} =1$ for the $0$-th iteration, and iteratively computing~\eqref{eq:de_cn_met},\eqref{eq:de_vn_met},\eqref{eq:de_vn_avg_met} it is possible to derive the probability that the messages passed along the edges of the graph are erasures as the number of iterations $\itnumber$ grows.

Let us define by $\app_j^{(\itnumber)}$ the probability that a data node of type $j$ is erased at the end of the $\itnumber$-th iteration. We have, 
\begin{align} \label{eq:de_app_met}
     \app_j^{(\itnumber)} &= \prod_{i=1}^{\cntypes} \left( \pcn{\itnumber} \right) ^{\a_{i,j}}.
\end{align}
The load threshold of a \ac{MET} \ac{IBLT} ensemble can be formally defined in terms of $\app_j^{(\itnumber)}$ as follows.
\begin{definition}[Load threshold] \label{def:load_threshold}
The load threshold $\eff^{\star}$ is the largest value of $\eff$ such that $\app_j^{(\itnumber)}\rightarrow 0$ for all $j\in \{1,2,\ldots, \dntypes\}$ when $m\rightarrow \infty$ and $\itnumber \rightarrow \infty$.
\end{definition}
Intuitively, in the regime of  $m\rightarrow \infty$, the probability that a random key-value pair is successfully recovered from the \ac{IBLT} tends to one, as long as the number $n$ of key-value pairs previously inserted in the \ac{IBLT} is below $\eff^{\star} m$.

\subsection{Numerical Examples}\label{sec:anal_num_res}
We report load thresholds complemented by finite-length simulations of the probability $P_e$ that a randomly chosen key-value pair is not present in the output list of the recovery operation. For illustration purposes, we consider two different \ac{MET} \ac{IBLT} designs with different load thresholds $\eff^{\star}$. 

\begin{example}
Consider a design $E_1$ with $3$ different data and cell node types, $\dntypes=\cntypes=3$,
a probability vector $\bm{p}=\begin{bmatrix} 0.2& 0.2& 0.6\end{bmatrix}$, $\bm{m} = \begin{bmatrix}m/3& m/3&  m/3\end{bmatrix}$, 
 and  a degree matrix 
\[\matA=\begin{bmatrix}
     1  &  2  &  1\\
     2  &  1  &  1 \\
     1  &  2  &  1\end{bmatrix}.\] 
     We obtain  $\eff^{\star} = 0.815$.
\end{example}
\begin{example}
For design $E_2$, we have $\dntypes=4$, $\cntypes=2$, $\bm{p}=\begin{bmatrix}0.046& 0.427& 0.398& 0.129\end{bmatrix}$, $\bm{m} = \begin{bmatrix} m/2 & m/2\end{bmatrix}$, and 
\[\matA=\begin{bmatrix}
     6  & 3 & 1 & 4\\
     14  & 0 & 2 & 6\end{bmatrix}.\]
      We obtain  $\eff^{\star} = 0.935$.
\end{example}

\begin{figure}[t!]
    \centering
    \includegraphics[width=0.99\columnwidth]{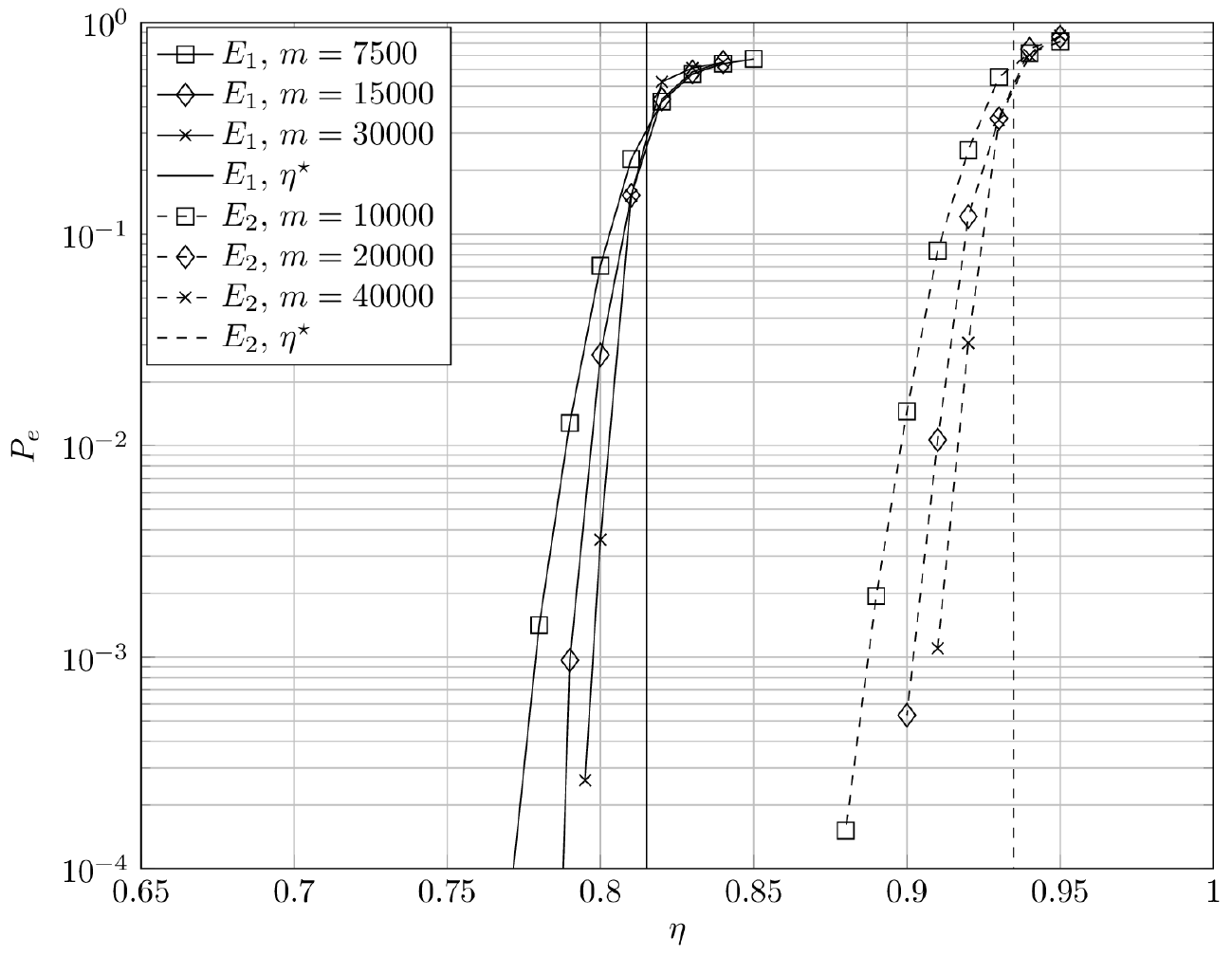}
    \caption{Probability of unsuccessful recovery of a  key-value pair, $P_e$, as a function of the channel load $\eff$ for two different \ac{MET} \ac{IBLT} designs and different values of $m$.}
    \label{fig:thres_sim}
\end{figure}

Figure~\ref{fig:thres_sim} depicts $P_e$ as a function of $\eff=n/m$ for  different $m$ considering both \ac{MET} \acp{IBLT} designs $E_1$ and $E_2$. Observe that for increasing $m$, $P_e$ shows a sharp drop at $\eff \approx \eff^{\star}$ where the load threshold $\eff^{\star}$ is indicated by a vertical line. Coarsely speaking, for large $m$ one can store approximately $\eff^{\star} m$ key-value pairs in the \ac{MET} \ac{IBLT}. The probability that the recovery of a randomly selected key-value pair fails goes to zero. Observe that design $E_2$ allows storing more key-value pairs than design $E_1$ in an \ac{IBLT} of a given size $m$ (for $m$ large enough). Note that design $E_2$ was obtained by running a computer search to improve the load threshold. In Section~\ref{sec:design} we will build on the analysis introduced earlier  in this section to find  good \ac{MET} designs suitable for a set reconciliation application.

	\section{Set Reconciliation Protocol} \label{sec:setrec_prot}

In this section, we explain how \acp{IBLT} can be used for set reconciliation. We describe a set reconciliation protocol  referred to as \protocol which relies on the presented \ac{MET} \acp{IBLT}.
\subsection{Set Reconciliation with \acp{IBLT}}

Following \cite{eppstein:2011}, let us first illustrate how \acp{IBLT} can be used for set reconciliation.
{First, hosts $A$ and $B$ insert all the elements of their sets $\SA$ and $\SB$ into the \acp{IBLT}  $\iblt{A}$ and  $\iblt{B}$, respectively. The mapping rules for both \acp{IBLT} $\iblt{A}$ and  $\iblt{B}$ are the same and are agreed upon beforehand by the hosts. Then, host $A$ sends $\iblt{A}$ to host $B$ that creates a so-called difference \ac{IBLT} $\iblt{\Delta}$. This is done by subtracting all the cells of  $\iblt{A}$ from those of $\iblt{B}$ applying Algorithm~\ref{alg:subtract} (which simply subtracts the \countfield field of the two cells, and xors the \payload field).}

We make the following observations regarding the difference \ac{IBLT}, $\iblt{\Delta}$:
 \begin{itemize}
 	\item The elements in $\SA \cap \SB$, the intersection of $\SA$ and $\SB$, are not present in the difference \ac{IBLT} $\iblt{\Delta}$ (it is as if they had been first inserted and then deleted). 
    \item { The elements in  the set difference, $\D = \SA \cup \SB \setminus \SA \cap \SB$} are still present in the difference \ac{IBLT}. In particular:
    \begin{itemize}
        \item  The elements in $\SB \setminus \{\SA \cap \SB\}$, those present in $\SB$ but not in $\SA$, have been inserted into $\iblt{\Delta}$ (and they have not been yet deleted).
 	      \item  The elements in $\SA \setminus \{\SA \cap \SB\}$, those present in $\SA$ but not in $\SB$, have been deleted from  $\iblt{\Delta}$, without having been inserted before.
    \end{itemize}
 	  \item {Thus, if we represent the difference \ac{IBLT} as a bipartite graph, we obtain a graph with $m$ cell nodes, and  $\delta = |\D|$ data nodes (only the elements of the set difference $\D$ are present). Furthermore, the degree and connectivity of the remaining data nodes is preserved compared to the bipartite graph of the \acp{IBLT} of host $A$ and/or $B$ (see Example~\ref{example:diff_iblt}). } 
 \end{itemize} 

\begin{example}[Difference \ac{IBLT}] \label{example:diff_iblt}

For ease of illustration, we consider an irregular \ac{IBLT} with $5$ cells in which  host $A$ has the set $\SA=\{\pair_1, \pair_2, \pair_3\}$, whereas host $B$ has $\SB=\{\pair_3, \pair_4 \}$, hence the set difference is $\Delta =\{ \pair_1,  \pair_2, \pair_4 \}$.
Figures~\ref{fig:protb_iblt_a} and \ref{fig:protb_iblt_b}, show the bipartite representation of $\iblt{A}$ and $\iblt{B}$ respectively, as well as the values stored in the cells of the \acp{IBLT}.
{The bipartite graph associated with $\iblt{\Delta}$, which is obtained by subtracting all  cells of $\iblt{A}$ from those of $\iblt{B}$, is shown in see Figure~\ref{fig:protb_iblt_Delta}}. Observe that there are no edges connected to the elements in the intersection of the sets, ${\SA \cap \SB} =\{ \pair_3\}$.

For illustration purposes, solid edges are attached to elements that are only in $\SB$ , i.e., in $\SB \setminus \SA \cap \SB$. 
Whenever a data node is connected to a cell node by a solid edge, the $\countfield$ field in the associated \ac{IBLT} cell will be increased by one. Dashed edges are attached to elements that are only in $\SA$ , i.e., in $\SA \setminus \SA \cap \SB$. 
Whenever a data node is connected to a cell node using a dashed edge, the $\countfield$ field in the associated IBLT cell will be decreased by one. By counting the number of solid and dashed edges we can derive the $\countfield$ field associated with each cell in $\iblt{\Delta}$.
Note however that  the \countfield of the cells in $\iblt{\Delta}$ no longer coincides with the degree of the associated cell nodes. For example, observing Figure~\ref{fig:protb_iblt_Delta} we can see that in cell $\cell_3$ we have $\countfield=0$, although actually, its associated cell node has degree $2$, i.e., we have $\payload = \pair_2 \oplus \pair_4$.

\begin{figure}[t!]
		\centering		
		\subfloat[$\iblt{A}$]{
			\includegraphics[width=0.8\columnwidth]{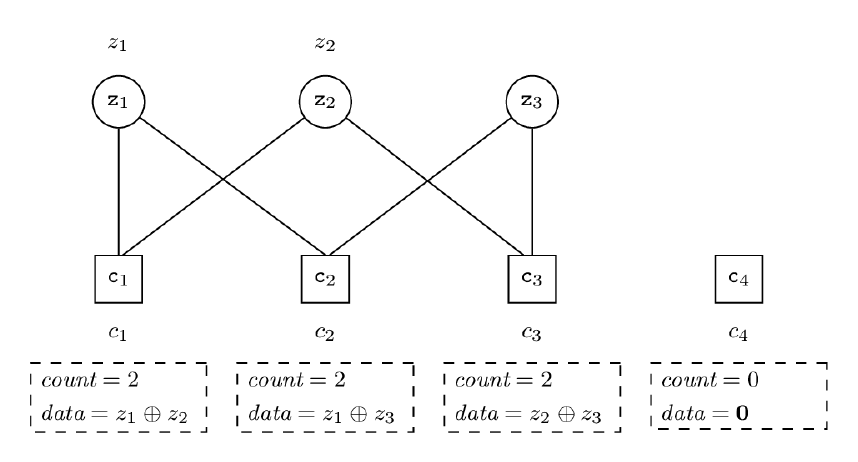}		
			\label{fig:protb_iblt_a}
		}		
  
		\subfloat[$\iblt{B}$]{
			\includegraphics[width=0.8\columnwidth]{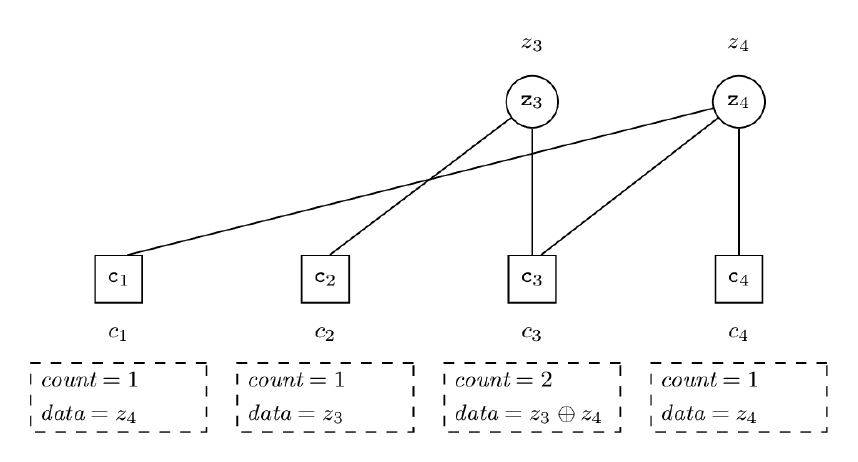}
			\label{fig:protb_iblt_b}
		}
		
		\subfloat[$\iblt{\Delta}$]{
			\includegraphics[width=0.8\columnwidth]{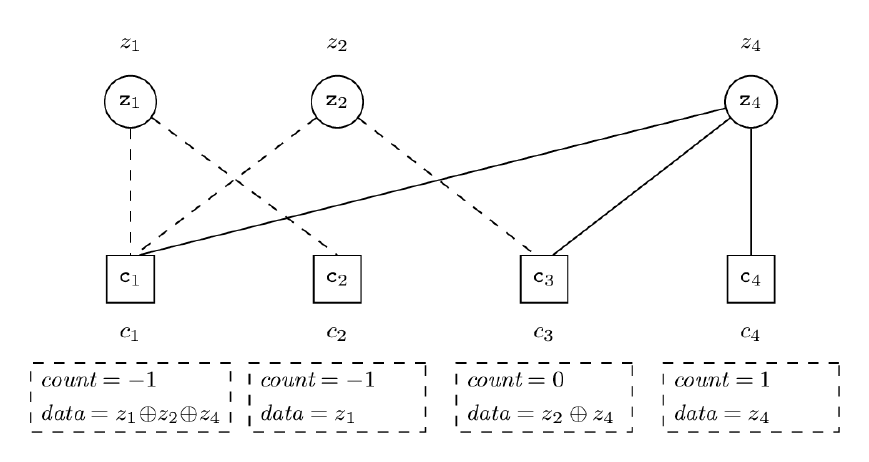}
			\label{fig:protb_iblt_Delta}
		}
		
		\caption{ Example of graph representation of the \acp{IBLT} generated locally by each of the hosts. The local \ac{IBLT} of host $A$ (transmitter), $\iblt{A}$, is shown in (a), and that of host $B$, $\iblt{B}$, is shown in (b). The difference \ac{IBLT} is illustrated in (c).}
		\label{fig:protoB_0}
\end{figure}
\end{example}

In order to be able to recover the set difference   $\D= (\SA \cup \SB) \setminus (\SA \cap \SB)$ from $\iblt{\Delta}$, the recovery algorithm must be altered  to deal with elements that have been deleted from the \ac{IBLT} without having been previously inserted. 
Let us recall that the standard recovery algorithm (Algorithm~\ref{alg:list}) searches for cells with $\countfield=1$, because in such cells a single set element has been inserted, and thus it can be directly recovered from the $\payload$ field. 
However, the modified recovery algorithm  (Algorithm~\ref{alg:listm}) needs also to account for elements that have been deleted from the IBLT without having been previously inserted (elements in $\SA \setminus \{\SA \cap \SB\}$).
In particular, the modified recovery algorithm has to search for cells where a single element has been \emph{inserted or deleted}. 
Following \cite{eppstein:2011}, we will refer to such cells as \emph{pure}. When a cell is pure, the $\payload$ field of the cell contains a copy of  the single element that has been inserted/deleted. 
\begin{example}[Purity]
In Figure~\ref{fig:protb_iblt_Delta}, we can see that cell $\cell_1$ has \countfield=-1, but it is not pure, since 3 key-value pairs have been mapped into it, $\pair_1$, $\pair_2$, and $\pair_4$. By contrast, cells $\cell_2$ and $\cell_4$ which have respectively \countfield -1 and 1 are pure since only one key-value pair has been mapped into them.
\end{example}

An issue consists in detecting pure cells, keeping in mind that the decoder only has access to the \ac{IBLT} cells and the mapping rules, but the graph structure is \emph{hidden}.
It is easy to see how $\countfield=\pm1$ is a necessary but not sufficient condition for a cell to be pure. Thus, altering the recovery algorithm to search for  cells with $\countfield=\pm 1$ is not sufficient. To  distinguish pure cells, one can rely on the fact that the keys are obtained from the values by using a function $\hashx(\cdot)$, \cite{eppstein:2011}. Hence, in addition to checking whether $\countfield= \pm1$, we also check if the key $z.x$ is valid, i.e., whether $z.x = \hashx(z.y)$ (see Algorithm~\ref{alg:checkp}).\footnote{In fact the condition $z.x = \hashx(z.y)$ renders the count field in the cells unnecessary for set reconciliation applications. Still, we keep it in our discussion for the  sake of clarity.}
Note however that the function $\hashx(\cdot)$ is a many-to-one mapping. Thus, this purity check may return true for an impure cell (see Section~\ref{sec:fail}).

Relying on this probabilistic purity check, it is possible to define a modified recovery algorithm (see Algorithm~\ref{alg:listm}) to recover the set difference from the difference \ac{IBLT} $\iblt{\Delta}$. This algorithm searches for pure cells. Whenever a pure cell is found, the only element mapped in the cell is added to the output list, and then this element is removed from the \ac{IBLT}. This is achieved by either deleting it from the \ac{IBLT} in case we have $\countfield = 1$ or adding it to the \ac{IBLT} in case we have $\countfield = -1$.

\begin{example}[Modified Recovery]
Let us consider modified recovery applied to the difference \ac{IBLT} shown in Figure~\ref{fig:protb_iblt_Delta}. For simplicity, we assume here that the purity check is perfect. Under this assumption, cell $\cell_2$, which has \countfield=-1 can be determined to be pure, allowing to recover $\pair_1$, which is removed from the \ac{IBLT} by relying on the addition operation. Similarly, cell $\cell_4$ is also found to be pure. This allows recovering $\pair_4$, which is removed from the \ac{IBLT} by calling the delete operation since we had \countfield=1 in $\cell_4$. Finally, $\pair_2$ can be recovered from either $\cell_1$ or $\cell_3$.
\end{example}

Fig.~\ref{fig:seq_diag}a shows the sequence diagram of difference digest, the IBLT based set reconciliation protocol proposed in \cite{eppstein:2011}. For simplicity, we have assumed here that the  size of the set difference $\d$ is known a priori. As it can be observed, after a request from node B, node A transmits its complete IBLT $\iblt{A}$. After receiving  $\iblt{A}$, node B computes the difference IBLT. Next, the modified recovery operation is used to attempt reconciling the sets. Finally, node B sends a (negative) acknowledgement to node A to communicate whether the reconciliation was successful or not.

\begin{figure}
	\centering
		\scalebox{0.9}{
	\begin{minipage}{0.49\textwidth}
		\vspace{-4.75mm}
		\begin{algorithm}[H]
			\small
			\caption{Cell Subtraction }\label{alg:subtract}
			\begin{algorithmic}
				\vspace{-1.5mm}
				\Procedure{$\cell^\dagger=$  Subtract($\cell$, $\cell^\prime$) }{}
				\State $\cell^\dagger$.\countfield =$\cell$.\countfield -$\cell^\prime$.\countfield
				\State $\cell^\dagger$.\payload =\xor($\cell$.\payload ,$\cell^\prime$.\payload )
				\EndProcedure
			\end{algorithmic}
		\end{algorithm}
	\end{minipage}
	}
	\scalebox{0.9}{
\begin{minipage}{0.49\textwidth}
	\begin{algorithm}[H]
		\small
		\caption{Check Purity}\label{alg:checkp}
		\begin{algorithmic}
			\vspace{-1.5mm}
			\Procedure{$b$=IsPure$(z)$}{}			
			\If{$z.x = \hashx(z.y)$}
			\State $b=\text{true}$
			\Else
			\State $b=\text{false}$
			\EndIf
			\EndProcedure
		\end{algorithmic}
	\end{algorithm}
\end{minipage}
}

\scalebox{0.9}{
\begin{minipage}{0.5\textwidth}
	\begin{algorithm}[H]
		\small
		\caption{Modified Recovery}\label{alg:listm}
		\begin{algorithmic}
			\vspace{-1.5mm}
			\Procedure{Recover$()$}{}			
			\While {$\exists i \in [1, m] \mid  \cell_{i}.\countfield = \pm 1~\&~ \ispure(\cell_{i}.\payload)= \text{true}$} 
			\State add $\pair=\cell_{i}.\payload$ to the output list
			\If{$\cell_{i}.\countfield =  1$}
			\State call Delete $(z)$
			\Else
			\State call Insert $(z)$
			\EndIf
			\EndWhile
			\EndProcedure
		\end{algorithmic}
	\end{algorithm}
\end{minipage}
}
\end{figure}

\subsection{Rate-Compatible Set Reconciliation Protocol}\label{sec:prot_def}
We propose a novel \protocol that relies on \ac{MET} \acp{IBLT}.
Consider a set reconciliation setting with two hosts, $A$ and $B$, which possess  sets $\SA$ and $\SB$, respectively. The hosts communicate with each other through a data network to reconcile their sets, and it is assumed that the protocol parameters, i.e., the parameters that define the \acp{IBLT} known to both parties. The protocol follows the following steps:

\begin{enumerate}
	\item Hosts $A$ and $B$ initialize to zero a \ac{MET} \ac{IBLT} of length $m$ cells, with $\cntypes$ different cell types and $\dntypes$ key-value pair types.
	Next, each host inserts all the elements of its set into the \ac{MET} \ac{IBLT}.
	The thus obtained \acp{IBLT} of $A$ and $B$ will be denoted $\iblt{A}$ and $\iblt{B}$, respectively. 
	We assume that the cells of $\iblt{A}$ and $\iblt{B}$ are ordered according to their type so that the first $m_1$ cells correspond to cells of type 1, followed by $m_2$ cells of type 2, and so on.

	\item Host $A$ starts sending the cells of its \ac{IBLT} $\iblt{A}$ one by one, starting with cell $\cell_1$, followed by $\cell_2$, and so on. Host $B$ creates an empty array of \ac{IBLT} cells $\iblt{\Delta_0}$. 
	\item When receiving the $i$-th cell of $\iblt{A}$, host $B$ subtracts it from the $i$-th cell of its local \ac{IBLT} $\iblt{B}$ (see Algorithm~\ref{alg:subtract}),   and it appends the  resulting cell to the difference \ac{IBLT}  $\iblt{\Delta_{i-1}}$ to obtain $\iblt{\Delta_i}$. The cell subtraction operation consists of subtracting the $\countfield$ fields of the cells and \xor-ing the $\payload$ fields.
	
	\item Host $B$ runs a modified recovery algorithm on $\iblt{\Delta_i}$ which, if successful, yields at its output the set difference $\D$ between $\SA$ and $\SB$ (see Algorithm~\ref{alg:listm}).
	
	\item If set reconciliation succeeds, host $B$ sends an acknowledgment to host $A$. 
	Otherwise, host $B$ does not send any acknowledgment, and host $A$ continues sending \ac{IBLT} cells. 
	
	\item If recovery is still unsuccessful after all $m$ cells have been transmitted, a failure is declared. 
\end{enumerate}

\begin{figure}
   \centering
   \subfloat[Difference digest]{   
   \includegraphics[width=7.5cm]{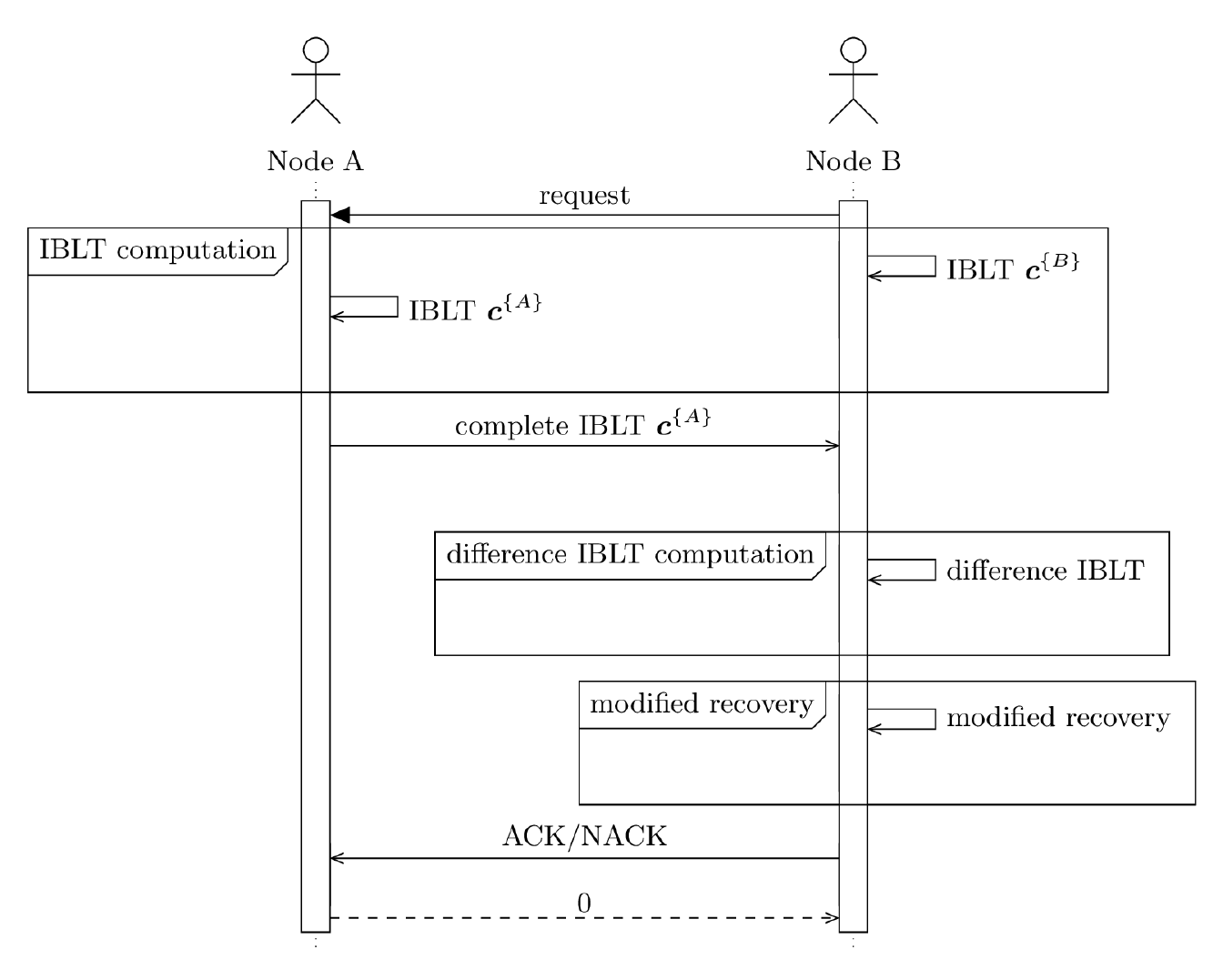}   
    }
    
    \subfloat[Rate-compatible set reconciliation]{  
    \includegraphics[width=8.6cm]{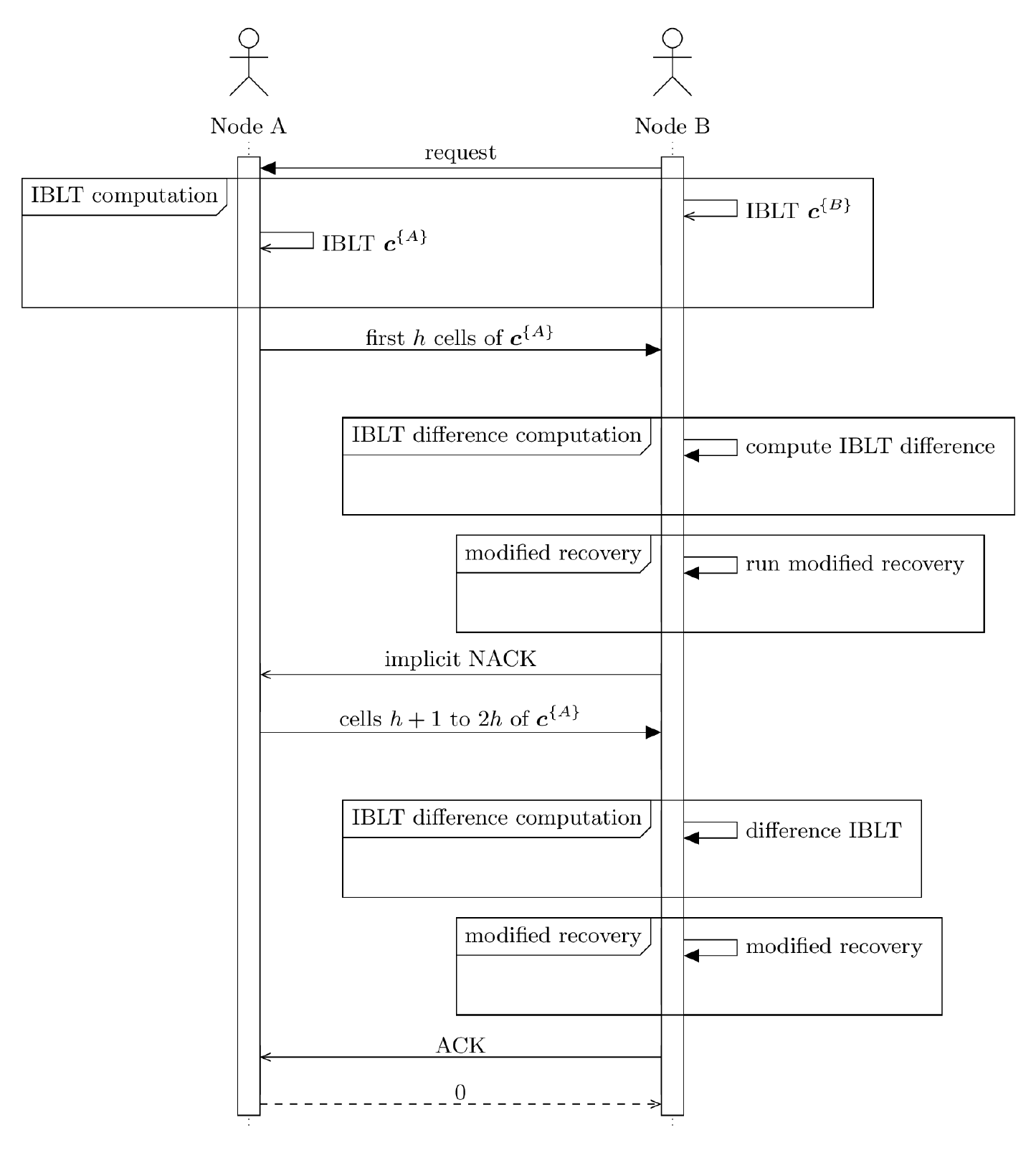}
    }
   \caption{Sequence diagrams of difference digest and the proposed \protocol\hspace{-1.3mm}. For simplicity, in difference digest we assume that the size of the set difference $\d$ is known a priori. In the rate-compatible protocol, we assume that cells are sent in packets (or groups) of $h$ cells. }
   \label{fig:seq_diag}
\end{figure}

Fig.~\ref{fig:seq_diag}b shows the sequence diagram of the proposed \protocol \hspace{-1.5mm}. In this sequence diagram assume that \ac{IBLT} cells are not sent one by one, but in groups of $h$ cells. After having received the first $h$ cells, node B attempts to renconcile the sets by running the modified recovery operation, which fails. This leads to the transmission of an \emph{implicit NACK}. Here, we say implicit because, in practice, rather than sending a NACK, node A could simply assume that set reconciliation did not yet succeed whenever no acknowledgement is received after a predefined amount of time. Next, Node A sends additional IBLT cells to node B. Node B then re-attempts set reconciliation by running the modified recovery operation, which now succeeds. Finally, Node B sends an acknowledgement to Node A and the protocol ends.
We make the following observations. First, when the protocol succeeds, only host $B$ is aware of the set difference. This is a common assumption in set reconciliation protocols. If necessary, host $B$ can then transmit a message to host $A$ to convey the set difference. 
Second, the protocol will fail if  recovery is still unsuccessful after having transmitted all $m$ cells. However, as illustrated in Section~\ref{sec:design}, one of the advantages of relying on \ac{MET} \acp{IBLT} is that $m$ may grow on-demand, allowing in principle to generate an unlimited number of cells so that reconciliation is eventually successful.

\begin{example} [Protocol] \label{example:prot}

{Figures~\ref{fig:protb_iblt_a_ratecomp} and \ref{fig:protb_iblt_b_ratecomp} show the graphical representation of the local \acp{IBLT} of host $A$ and $B$ ($\iblt{A}$ and $\iblt{B}$), respectively. Both bipartite graphs have a total of ${m=4}$ \ac{IBLT} cell nodes, out of which 2 are of type 1 and 2 of type 2, i.e., $m_1=m_2=2$. Furthermore, we have $\matA^T=\begin{bmatrix} 1 &1 \end{bmatrix}$, i.e., there is a single data node type. 
We have $\SA=\{ \pair_1, \pair_2, \pair_3 \}$ and $\SB=\{ \pair_2, \pair_3, \pair_4 \}$. Thus we have that the set difference is $\D=\{ \pair_1, \pair_4 \}$.}

According to the rate-compatible protocol, host $A$  sends its local \ac{IBLT} cells one by one to host $B$. After receiving the $i$-th cell, host $B$ subtracts it from the $i$-th cell of its local IBLT. The thus obtained difference \ac{IBLT} is denoted $\iblt{\Delta_i}$. 
Thus, for $h\leq i$,  the $h$-th cell of  $\iblt{\Delta_i}$ is obtained by subtracting the $h$-th cell of  $\iblt{A}$ from the $h$-th cell of $\iblt{B}$. Moreover, we have no information about the $h$-th cell of $\iblt{\Delta_i}$ for $h > i$. These cells are considered erased.
After that, host $B$ attempts to recover the set difference by running the modified recovery algorithm, which amounts to a peeling operation on the graph representation of $\iblt{\Delta_i}$. Figures~\ref{fig:protb_iblt_delta_1}, \ref{fig:protb_iblt_delta_2}, and \ref{fig:protb_iblt_delta_3} show the graph representation of $\iblt{\Delta_i}$, for $i \in \{1,2,3\}$. {Note that the bipartite graph is provided for illustration purposes only. In the beginning, it is hidden to host $B$ and it is successively revealed during the recovery operation.} In particular, host $B$ attempts set reconciliation by inverting $\iblt{\Delta_i}$, i.e., by carrying out the modified recovery operation (Algorithm~\ref{alg:listm}), relying only on the $i$ cells it has so far received, and treating \ac{IBLT} cells $i+1$ onwards as if they had been erased. For this reason, when representing $\iblt{\Delta_i}$ we mark cell nodes $\cellnode_{i+1}, \cellnode_{i+2}, \dots$ in gray.

Observe that reconciliation is not successful after host $B$ has received the first \ac{IBLT} cell since $\cellnode_1$ is empty in $\iblt{\Delta}$ (and all other cells are still erased).
Similarly, after having received 2 cells, reconciliation is still unsuccessful, since $\cellnode_2$ is not pure.
Finally, reconciliation is successful after having received 3 cells. In particular, cell $\cellnode_3$ is pure, allowing to recover $\pairnode_1$, which is then removed from the \ac{IBLT}. This renders cell $\cellnode_2$ pure, allowing to recover $\pairnode_4$.
Since set reconciliation is successful on $\iblt{\Delta_3}$, host $B$ sends an acknowledgment to host $A$, and the reconciliation protocol terminates. Cell $\cellnode_4$ is in this case never transmitted.

	\begin{figure}[t!]
	\centering
			\subfloat[$\iblt{A}$]{
				\includegraphics[width=0.45\columnwidth]{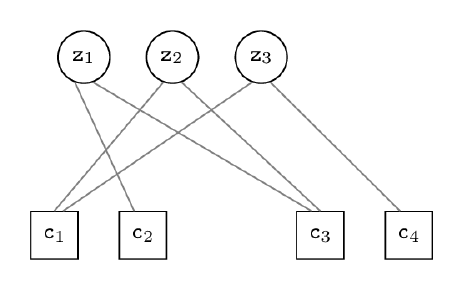}
				\label{fig:protb_iblt_a_ratecomp}
			}		
			\subfloat[$\iblt{B}$]{
				\includegraphics[width=0.45\columnwidth]{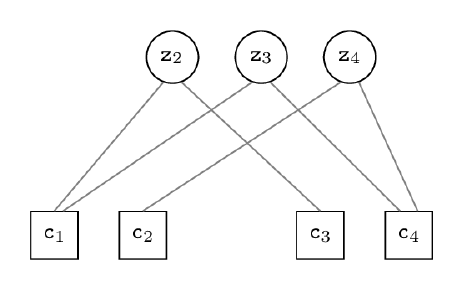}
				\label{fig:protb_iblt_b_ratecomp}
			}

			\subfloat[$\iblt{\Delta_1}$]{
				\includegraphics[width=0.45\columnwidth]{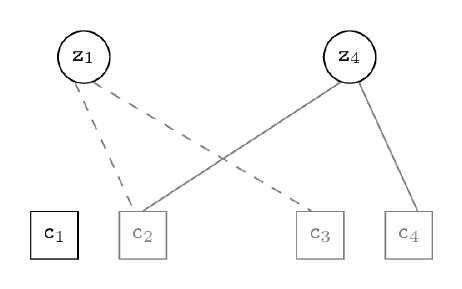}
				\label{fig:protb_iblt_delta_1}
			}   		
			\subfloat[$\iblt{\Delta_2}$]{
				\includegraphics[width=0.45\columnwidth]{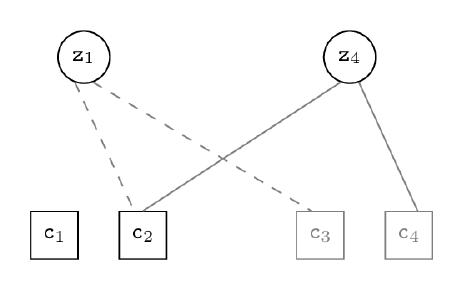}
				\label{fig:protb_iblt_delta_2}
			}	
   
        \begin{center}   			
			\subfloat[$\iblt{\Delta_3}$]{
				\includegraphics[width=0.45\columnwidth]{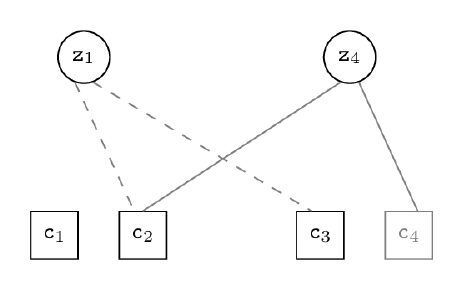}
				\label{fig:protb_iblt_delta_3}
			}	
		\end{center}
		
		\caption{Set reconciliation example. Figures (a) and (b) show  the graph representation of $\iblt{A}$ and $\iblt{B}$. Figures (c), (d), and (e) show the graph representation of  $\iblt{\Delta_i}$ , $i=1,2,3$, i.e., after having node $B$ has received 1, 2 and 3 cells respectively. The cell nodes shown in gray represent \ac{IBLT} cells which have not yet been received and are thus treated as if they were erased (or punctured).}
		\label{fig:protoB}
	\end{figure}

\end{example}

{The set reconciliation protocol described above has the advantage of exhibiting a low complexity since it relies on the modified recovery operation, which is equivalent to a peeling operation applied to the graph representation of the difference \ac{IBLT}. One could argue that this is suboptimal since host $B$ is not making use of his knowledge of $\SB$. In fact, it is possible to improve the performance of the modified recovery operation (peeling) by relying on the knowledge of $\SB$. For example, whenever the modified recovery operation (peeling) gets stuck because there are no pure cells, one could try to delete different elements of $\SB$ with the hope that they are in $\Delta$ and that their deletion generates some extra pure cells allowing modified recovery (peeling). Such an approach was proposed in \cite{gabrys2014set} and it was shown that it can improve performance at the cost of an increased complexity.
}

\subsection{Reconciliation Failures}\label{sec:fail}

In the protocol described above, two types of events can lead to undetected failures. 
The first type of failure is associated with the purity check (see Algorithm~\ref{alg:checkp}) returning true for a cell that is not pure.
When this happens, an \emph{erroneous} key-value pair will be added to the output list of the recovery algorithm.
Furthermore, the removal of this erroneous key-value pair from the \ac{IBLT} will corrupt a number of cells. 
With high probability, these corrupted cells will lead the receiver (host $B$) to believe that there are still some key-value pairs that have not been recovered, and the set reconciliation will run until all the $m$ cells have been transmitted, declaring then a reconciliation failure. 
Under the assumption that $\hashx(x)$ returns a random number between 1 and $2^\keylength$, we have that the purity check of an impure cell fails with probability $2^{-\keylength}$.
Assuming a total of $m$ cells, a set difference size $\d$, and an average data node degree $\bar d$, the number of different purity checks that have to be carried out on impure cells can be upper bounded as $m+ \d \bar d$. Thus, we have that the probability of this type of failure can be upper bounded by $ (m+ \d \bar d) 2^{-\keylength}$. Hence, by choosing $\keylength$ to be large enough, this probability can be made arbitrarily small. 

The other source of failure of the protocol is the fact that host $B$ may mistakenly assume that all key-value pairs have been recovered, while this is not the case. This happens when, after having received $i$ cells, the modified recovery (peeling decoding) of $\iblt{\Delta_i}$  leads to an empty \ac{IBLT}, however, there are one or more key-value pairs that have not yet been recovered.
These yet unrecovered key-value pairs are not mapped to any of the first  $i$ cells, i.e., they have no edges to any of the first $i$ cell nodes. Thus, there is no way for $B$ to detect the existence of these unrecovered key-value pairs. 
The best way of dealing with this second type of failure is to never terminate the protocol before having received at least $m_1$ cells, i.e., all the cells of type $1$. Assuming that $\a_{1,j}>0 \, \forall j$, this effectively solves the problem since in this case all key-value pairs will be mapped at least to one of the first $m_1$ cells.

\subsection{Protocol Variants}\label{sec:implementation}

From a practical viewpoint, it might be of interest to change the insertion (encoding) and recovery (decoding) of \ac{MET} \acp{IBLT} in Section~\ref{sec:IBLT}. It is not necessary to generate all $m$ cells and to recover all $m$ cells at once. Rather, encoding and decoding can proceed in rounds similar to rate-compatible coding schemes (see e.g., \cite{HKM04}). We may define an encoding index vector $\begin{bmatrix} i_0=0 &i_1& i_2 & \ldots\end{bmatrix}$ of positive integers $0<i_{1}<i_2 \ldots$. In encoding round $\alpha$ the encoder may generate $i_{\alpha}-i_{\alpha-1}$ additional cell nodes. We may for instance choose $i_{\alpha}-i_{\alpha-1}=m_{\alpha}$. 
Likewise, we may define a decoding index vector $\begin{bmatrix} j_0=0 & j_1 & j_2 & \ldots \end{bmatrix}$ of non-zero integers $0<j_{1}<j_2 \ldots$. Once $j_{\alpha}$ cells are received, the decoder performs round $\alpha$ of decoding. If the decoding attempt fails it waits for additional $j_{\alpha+1}-j_{\alpha}$ cells and retries again. For instance, we may choose $j_{\alpha+1}-j_{\alpha}=1 ~\forall \alpha$, i.e., we attempt decoding once receiving a cell.

\section{MET IBLT Design}\label{sec:design}

The \protocol in Section~\ref{sec:setrec_prot} requires a suitable \ac{MET} \ac{IBLT} design that provides good performance for a wide range of  sizes of the set difference $\d$, considering that $\d$ is not (perfectly) known a priori. A simple solution would be to perform an \ac{IBLT} design for large $\d$  by relying on the results of Section~\ref{sec:analysis}. However, when $\d$ is small, there is no guarantee that this scheme will work with reasonable communication complexity, i.e., that it will be efficient.
To illustrate why this is the case, let us consider a $k$-regular\footnote{in a $k$-regular \ac{IBLT} all key-value pairs and cells are of the same type, and all key-value pairs are mapped exactly on $k$ cells. Thus, we have ${\dnp=p_1 =1}$, and $\matA=k$.} \ac{IBLT} of length $m$. We focus on the graphical representation of the difference \ac{IBLT} $\iblt{\Delta_{h}}$ after $h$ cells have been exchanged. Let us now consider a key-value pair node associated with the set difference $\D$. The probability that this node has no edges attached to any of the first $h$ cell nodes can be approximated as 
\begin{align}\label{eq:dec_failure}
  \left( \frac{m-h}{m} \right)^k.
\end{align}
Note that the expression in~\eqref{eq:dec_failure} yields a lower bound on the probability of recovery failure $P_e$. For fixed $k$, e.g., $k=3$, this probability is high, unless $h$ is  close to $m$ (see also \cite[Section~6]{Shokrollahi2009} for a more detailed discussion). 
Thus, we expect the \protocol to perform poorly when employing standard \acp{IBLT}. This is illustrated by simulation results in Section~\ref{sec:numres}. The issue can be overcome by  \ac{MET} \acp{IBLT}. In particular, we can choose $m_1$ to be small, and forbid the protocol to terminate before having exchanged $m_1$ cells. This ensures that every key-value pair in $\D$ is mapped to a cell of type $1$.

\subsection{Design Example} \label{sec:desing_example}

Consider a \ac{MET} \ac{IBLT} design in which the number of cells of type $i$ is $m_i= 2^{i-1} m_1$.
We use the tools in Section~\ref{sec:analysis} to predict the asymptotic performance of the \ac{MET} \ac{IBLT}.
For simplicity, we set $\dntypes=3$, and consider a degree matrix in the form
     \[ 
     \matA^\text{T}=
     \begin{bmatrix}
	\a_{1,1}  & \a_{2,1} & \a_{3,1} & \a_{4,1} & \a_{5,1} & \cdots & \a_{5,1} \\
	\a_{1,2}  & \a_{2,2} & \a_{3,2} & \a_{4,2} & \a_{5,2} & \cdots & \a_{5,2} \\
	\a_{1,3}  & \a_{2,3} & \a_{3,3} & \a_{4,3} & \a_{5,3} & \cdots & \a_{5,3} 
\end{bmatrix}
       \]
where the fifth and all successive rows of $\matA$ are equal to the fourth row. This yields a design in which a potentially unlimited number of cell types $\cntypes$ can be generated.\footnote{Strictly speaking, the analysis presented in Section~\ref{sec:analysis} only applies to the case in which the number of cell types $\cntypes$ is constant. In practice, this does not hinder us from allowing the number of cell types to grow if needed.}
Let us now denote by $\eff^{\star}_i$ be the load threshold of the \ac{MET} \ac{IBLT} when considering only the first $\sum_{j=1}^{i} m_j$ cells.
The parameters of this design,  $\bm p$ and $\matA$, are determined by relying on multi-target optimization. In particular, we choose  $\bm p$ and $\matA$, to maximize the minimum among $\eff^{\star}_i$,
$$
(\bm p , \matA )  = \underset{\bm p, \matA}{\arg \max} \left( \min_{i} \eff^{\star}_i \right) .
$$
An approximate solution to this optimization problem was obtained by relying on simulated annealing \cite{kirkpatrick1983optimization}, an optimization algorithm to approximate the global optimum of a given function. For the current design example, we set $ i\in \{1, 2, \dots, 8\}$. Additionally, we introduce the constraint $\a_{i,j}\leq 5\, \forall i,j$ in order to limit complexity.\footnote{By removing the constraint our optimization algorithm yields only slightly higher thresholds.} The optimization yields

\[
\bm p=\begin{bmatrix} 0.1959 & 0.1904& 0.6137\end{bmatrix} \]
and 
\[
\matA^\text{T}=\begin{bmatrix}
	3  & 1 & 1 & 1 & 1 & \cdots & 1 \\
	4  & 4 & 4 & 4 & 5 & \cdots & 5 \\
	2  & 1 & 1 & 1 & 1 & \cdots & 1 
\end{bmatrix}
     \]
and the load thresholds obtained are
\begin{align}
\bm{\eta}^{\star}= [ & 0.7948,    0.7837,    0.7882,    0.8025 \\
            &   0.8042,    0.7967,    0.7895, 0.7856 ]. 
\end{align}
Although we only consider the first $8$ cell types in our optimization, the resulting \ac{MET} \ac{IBLT} still retains a good performance when additional cell types are included by replicating the last row $\matA$. In particular, we have $\eta^{\star}_9=0.7842$, $\eta^{\star}_{10}=0.7837 $ and $\eta^{\star}_{i}= 0.7830$ for $i\geq 11$.

\begin{remark}
This multi-target optimization  only  considers the performance at intermediate points in which all the cells of type $i$ have been received. In the Appendix, we introduce an extension of the analysis in Section~\ref{sec:analysis} that allows deriving  the load threshold of a \ac{MET} \ac{IBLT} when  only a fraction of cells of a certain type has been received.
\end{remark}

\subsection{Simulation Setup}
We consider a setup with two hosts, $A$ and $B$, who want to reconcile their sets $\SA$ and $\SB$ by communicating over a data network without having any knowledge about the  size of the set difference $\d$.
We apply the set reconciliation protocol of Section~\ref{sec:prot_def} with a \ac{MET} \ac{IBLT} from Section~\ref{sec:design} and compare the performance with two other schemes available in the literature, difference digest \cite{eppstein:2011} and \acf{CPI} \cite{Minsky2003}.

Difference digest \cite{eppstein:2011} is an \ac{IBLT}-based set reconciliation algorithm that relies on the use of regular \acp{IBLT}. In contrast to the other two schemes, difference digest requires an estimate of the set difference size, which is computed in an initial communication round, in which host $B$ sends to host $A$ a data structure known as strata estimator which is used to obtain an estimate $\dest$ of the set difference size $\d$. 
Host $A$ then calculates a worst-case set difference size $t=c\cdot \dest$, where $c$ is chosen so that $t>\d$ with high probability (e.g., $0.99$). Next, host $A$ creates an \ac{IBLT} of size $m$, where $m$ is chosen so that the recovery operation is successful with high probability (e.g., probability $0.99$) when $t$ key-value pairs are inserted  in the \ac{IBLT}. After that, host $A$ inserts its set $\SA$ in the \ac{IBLT} $\iblt{A}$ and sends it to host $B$. 
Host $B$ then also inserts his set $\SB$ in an \ac{IBLT} $\iblt{B}$ of size $m$ and then computes the difference \ac{IBLT} by subtracting the cells of $\iblt{B}$ and $\iblt{A}$, which yields the \ac{IBLT} $\iblt{\Delta}$ that also has length $m$. Next, host $B$ attempts to recover the set difference $\D$ by applying a  modified recovery operation (see Algorithm~\ref{alg:listm}). If recovery is successful, set reconciliation succeeds, otherwise, we will simply assume that set reconciliation is declared as unsuccessful.\footnote{In \cite{eppstein:2011} it was actually proposed to reattempt reconciliation by employing \acp{IBLT} of length $2m$. By allowing difference digest to fail we will slightly penalize the comparison with the two other schemes which continue until reconciliation is successful.}

The second scheme, \ac{CPI}, makes use of the characteristic polynomial of a set. Consider a set  $\S=\{ \pair_1, \pair_2, ..., \pair_n \}$ where the elements $\pair_i$ are chosen from $\univ=\{0,1\}^\ell$. The characteristic polynomial of set $\S$ over the finite field $\mathsf{F}_q$ with $q>2^{\ell}$ is defined as
\[
\charpol_{\S}(Z) = (Z-\pair_1) (Z-\pair_2) \dots (Z-\pair_n)
\]
where one relies on an {injective} mapping of elements of $\univ$ to $\mathsf{F}_q$.
The key principle behind \ac{CPI} is that given $n$ evaluations of $\charpol_{\S}(Z)$ it is possible to recover the set $\S$ since its elements correspond to the zeros of the characteristic polynomial  $\charpol_{\S}(Z)$. 
As we sketch next, \ac{CPI} can be also operated in a rateless fashion \cite{Minsky2003} to minimize the communication cost and does not require an estimate of the set difference size.
First of all, host $A$ chooses a list of evaluation points  $\w_1, \w_2, \dots$ at which it will evaluate the characteristic polynomial of its set.\footnote{in particular, the evaluations points $\w_1, \w_2, \dots$ are chosen among those elements of $\mathsf{F}_q$ to which no element of $\univ$ is mapped.}
Next, host $A$ continuously evaluates the characteristic polynomial of his set $\S_A$, $\charpol_{\S_A}(Z)$, at  the  evaluation points, $\w_1, \w_2, \dots$ and sends the evaluations $\charpol_{\S_A}(\w_1)$, $\charpol_{\S_A}(\w_2),\dots$ to host $B$, and it does so until it receives an acknowledgment from host $B$ which indicates that reconciliation was successful. Thereby, host $B$ continuously attempts to recover the set difference $\D$, by evaluating the ratio of two characteristic polynomials ${ \charpol_{\S_A}(\w_i) } / { \charpol_{\S_B}(\w_i)}$. 
For aset difference size of $\d$ this protocol has overall complexity $\mathcal{O}(d^4)$  \cite{Minsky2003}. 

\subsection{Simulation Results}\label{sec:numres}
We compare the performance of the three schemes in terms of communication cost, i.e., in terms of the amount of data, measured in bits, that needs to be exchanged.
For illustration purposes, we consider the setup from \cite{eppstein:2011} in which the cardinality of the  $\SA$ and $\SB$ is $10^5$, and the set elements are chosen from $\univ=\{0,1\}^{32}$. Following \cite{eppstein:2011}, we allocate a total of $12$ bytes for each \ac{IBLT} cell: $4$ bytes for the $\countfield$ field and $8$ bytes for the $\payload$ field ($4$ bytes for the key and $4$ bytes for the value). 
For difference digest we use the same configuration as in \cite{eppstein:2011}, thus in the initial communication round to estimate the set difference size, we employ a hybrid strata estimator with 7 strata, each with 80 cells per \ac{IBLT}, and a min-wise estimator with 2160 hashes. Hence, the resulting strata estimator (the data structure used to estimate the set difference size) has a total size of $15$ kB.
In the second communication round in which the actual set reconciliation takes place, we use a degree-$3$ regular \ac{IBLT}.
For \ac{CPI}, we use a configuration in which every evaluation point takes $10$ bytes.\footnote{We operate on a finite field with $2^{40}$. Hence, we need $5$ bytes to specify the point at which we evaluate the polynomial, and another $5$ bytes for the evaluation of the polynomial.}  
Finally, for the \protocol we  use $12$ bytes for each \ac{IBLT} cell, the same as in difference digest, and the \ac{MET} \ac{IBLT} parameters reported in Section~\ref{sec:design}.
In all cases, we assume that the protocol configuration is known a priori to both reconciliation parties and does not need to be signaled. For example, for the rate-compatible protocol, this assumption translates into the knowledge of the parameters $\bm p$, $\matA$, $\bm{m}$ as well as the hash functions $\hash$, $\hashtype$, and $\hashx$.

\begin{figure} [t!]
		\centering
		\includegraphics[width=0.99\columnwidth]{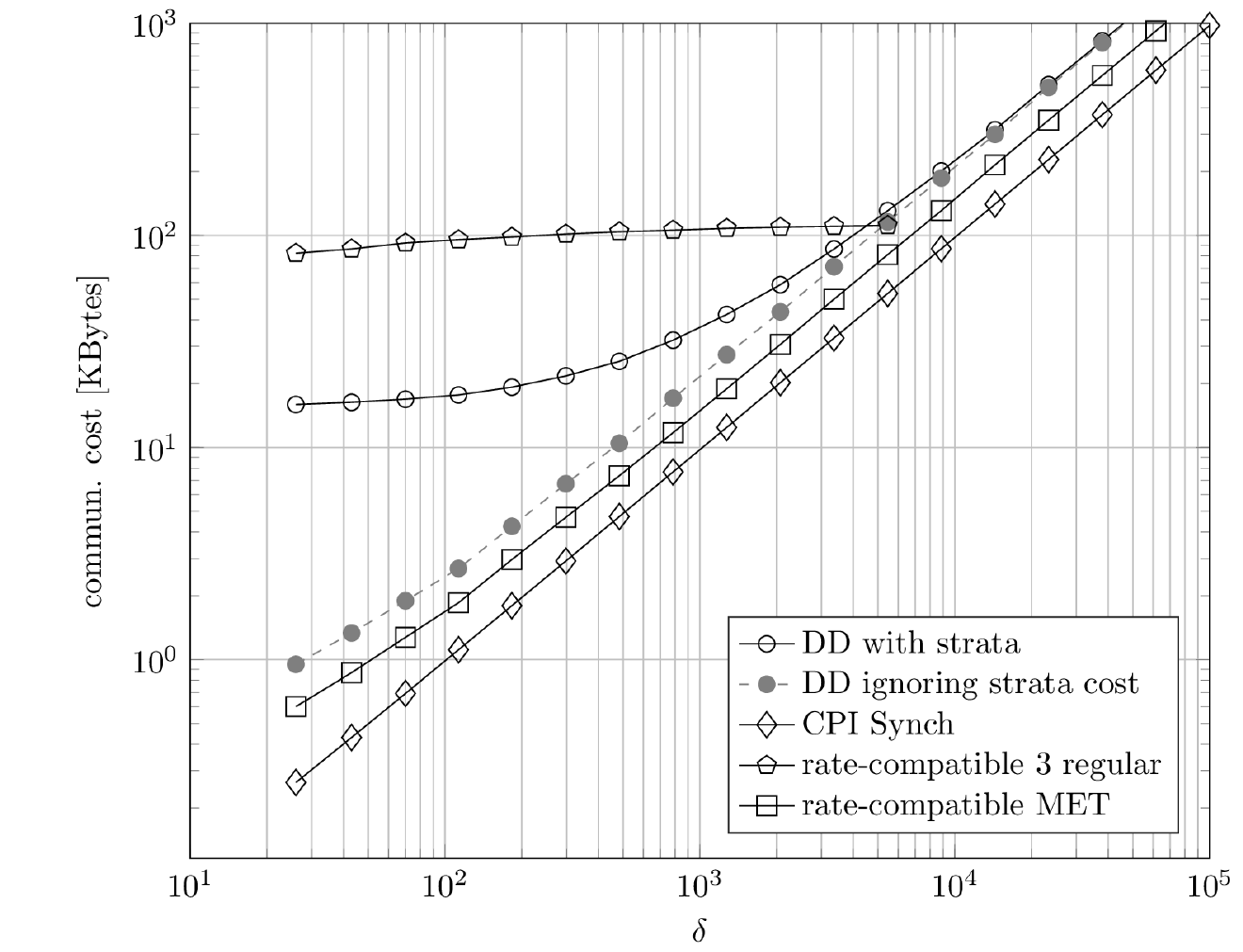}
	
		\caption{Data transmission required to reconcile sets with $10^5$ elements as a function of the  set difference size $\d$. The solid line with circle markers represents the data transmitted by difference digest \cite{eppstein:2011}, including the strata estimator. The gray dashed line with circle markers represents the data transmitted by difference digest ignoring the size of the strata estimator for the computation of the communication cost. 
		The solid line  with diamond markers represents the CPI scheme in \cite{Minsky2003}. Finally, the  solid line with square and pentagon markers represent the \protocol introduced in this paper using the \ac{MET} \ac{IBLT} form Section~\ref{sec:desing_example} and a 3-regular \ac{IBLT}.}
		\label{fig:comparison}
	\end{figure}

Figure~\ref{fig:comparison} shows the communication cost, i.e., the total amount of transmitted data needed for set reconciliation for the three schemes. For difference digest, the figure also reports the total data transmission without considering the strata estimator (which takes exactly $15$ kB in our example).
For the \protocol we show two performance curves. The first one uses the \ac{MET} \ac{IBLT} design described in Section~\ref{sec:design} setting $m_1=50$. The second one uses a 3-regular distribution with $m=10^4$.

Observe that the \protocol when using a \ac{MET} \ac{IBLT} has only a slightly higher communication cost than the more complex \ac{CPI} scheme, and it outperforms difference digest. 
This is remarkable since the complexity of our \protocol\!\!, when relying on the design introduced in Section~\ref{sec:design}, is $\mathcal{O}( \d \log(\d))$, whereas that of \ac{CPI} and difference digest (ignoring the strata estimator) are $\mathcal{O}(\d^4)$ and $\mathcal{O}(\d)$, respectively.
This can be attributed to two different effects. The first one is the presence of an initial communication round to estimate the set difference size in difference digest. This inefficiency dominates the performance for small $\d$ since in this case, the size of the strata estimator is much larger than the actual data that has to be exchanged to reconcile the sets.
The second one is the rate-compatible nature of the protocol which allows transmitting \ac{IBLT} cells incrementally. %

The performance of the \protocol when using a degree 3 regular \ac{IBLT} is also depicted. Observe that the performance is very poor compared to the \ac{MET} \ac{IBLT} design. This is  due to the presence of key-value pairs that are not mapped to the first cells exchanged (see discussion in Section~\ref{sec:design}) and hence cannot be recovered upon reception of those cells.

	\section{Discussion}\label{sec:conclusions}

In this paper, we have presented a  novel set reconciliation algorithm whose performance can approach that of \acf{CPI} but with lower algorithmic complexity.
The proposed scheme can be seen as an evolution of difference digest \cite{eppstein:2011} with two major modifications.
The first one is the concept (and analysis of) \ac{MET} \acp{IBLT}, a generalization of the \ac{IBLT} data structure, in which one allows for different types of \ac{IBLT} cells and  key-value pairs.
The second modification is the fountain-like protocol, meaning that the number of \ac{IBLT} cells to be exchanged is not fixed a priori. Instead, one can incrementally transmit as many cells as necessary, and new cells can in principle be generated on the fly.  

The novel scheme presents some advantages compared to difference digest. 
It is not necessary to estimate the cardinality of the set difference which usually requires an additional communication round and introduces a substantial overhead. Likewise, we only have to transmit as many cells as necessary, instead of having to oversize the \ac{IBLT} to cope with errors in the estimate of the cardinality of the set difference.

The proposed scheme combines the low complexity of difference digest and the efficiency of \ac{CPI}, and could thus represent an appealing solution to many data synchronization problems in  practical distributed systems.
A possible application is  the reconciliation of large databases. A  concrete example is mempool synchronization in the Bitcoin network \cite{bovskov2022gensync}. 
Another important application is remote file synchronization \cite{agarwal2006bandwidth, yan2008algorithms, Gentili2015}. 

{There are several of interesting extensions to this work. 
An example is the extension of the proposed scheme to a multi-party set reconciliation setting, as proposed for standard \acp{IBLT} in \cite{MitzenmacherP18}. 
Another example concerns a broadcasting setting in which a parent node holds a set $\S_P$ and multiple child nodes hold sets $\S_{C_1}$, $\S_{C_2}$, etc. In this setting, the set difference between $\S_{C_i}$ and  $\S_P$ can be different for different $i$. The goal could be for the parent node to convey its set $\S_P$ efficiently to the children nodes by transmitting \ac{IBLT} cells over a broadcast channel. By relying on the proposed rate flexible constructions, a child node with a small set difference with respect to the parent node would obtain $\S_P$ faster than a child node with a larger set difference.
This extension could have applications in settings in which a parent node maintains a database 
and multiple child nodes maintain a local copy of the database.
}

	\section*{Acknowledgements}
	The authors would like to thank Federico Clazzer for providing some software routines that were used for the Monte Carlo simulations. 	
	
	The research leading to these results has been carried out under the framework of the project ”SiNaKoL”. The project started in January 2022 and is led by the Program Coordination Security Research within the German Aerospace Center (DLR), whose support we greatly appreciate.


	\appendix
	
For the \protocol\!, cell nodes are transmitted one by one. The recovery operation can be attempted anytime even if not all cells of a type have been transmitted and hence received. We extend the analysis of Section~\ref{sec:analysis} to this case. The following analysis is also of interest in case a non-ideal communication channel between transmitter and receiver is assumed which introduces losses of cells (erasures). 

Let us denote by $\epsilon_i$ the probability that a cell node of type $i$ is not received (erased). For $m_i\rightarrow \infty$ (by the law of large numbers) $\epsilon_i$ is the fraction of erased cell nodes of type $i$. For the analysis, we assume that these cell nodes are part of the \acp{IBLT} bipartite graph, but the outgoing messages (i.e., probability of erasure) on all connected edges are one throughout all iterations. The remaining fraction of $(1-\epsilon_i)$ cell nodes of type $i$ will send out a message given by \eqref{eq:de_cn_met}. Overall, the average erasure probability from a cell node of type $i$ to a data node in case of erasures is modified as follows,
\begin{align} \label{eq:CN_update_erasure}
\pcn{\itnumber}=\epsilon_i+(1-\epsilon_i) \left(1-e^{ - \frac{\eta}  {\cnfraci} \avgdi  \pvnavg{\itnumber}}\right) .
\end{align}
Observe that by not transmitting cell nodes of a certain type, the load is increased, i.e.,
\begin{align} \label{eq:load_update_erasure}
\eta=\frac{n}{\sum_{i=1}^{\cntypes} (1-\epsilon_i) m_i}.
\end{align}
Apart from the modification in \eqref{eq:CN_update_erasure} and the adjustment of the load in \eqref{eq:load_update_erasure} the load threshold computation can be carried out as described in Section~\ref{sec:analysis}.

\end{document}